\documentclass[fleqn,usenatbib]{mnras}

\usepackage{newtxtext,newtxmath}

\usepackage[T1]{fontenc}

\DeclareRobustCommand{\VAN}[3]{#2}
\let\VANthebibliography\thebibliography
\def\thebibliography{\DeclareRobustCommand{\VAN}[3]{##3}\VANthebibliography}

\usepackage{graphicx}	
\usepackage{amsmath}	

\newcommand{\bigO}[1]{\mathcal{O}(#1)}

\title[TREVR2]{TREVR2: Illuminating fast $N\log_2\,N$ radiative transfer}

\author[J. W. Wadsley et al.]{
James W. Wadsley,$^{1}$\thanks{E-mail: wadsley@mcmaster.ca (JWW)}
Bernhard Baumschlager,$^{2}$
and Sijing Shen$^{2}$
\\
\\
$^{1}$Department of Physics and Astronomy, McMaster University, Hamilton, Ontario L8S 4M1, Canada\\
$^{2}$Institute of Theoretical Astrophysics, University of Oslo, P.O. Box 1029 Blindern, 0315 Oslo, Norway\\
}

\date{Accepted XXX. Received YYY; in original form ZZZ}

\pubyear{2023}

\begin{document}
\label{firstpage}
\pagerange{\pageref{firstpage}--\pageref{lastpage}}
\maketitle

\begin{abstract}
{
We present {\sc trevr2} (Tree-based REVerse Ray Tracing 2), a fast, general algorithm for computing the radiation field, suitable for both particle and mesh codes.
It is designed to self-consistently evolve chemistry 
for zoomed-in astrophysical simulations, such as cosmological galaxies with both internal
sources and prescribed background radiation, rather than large periodic volumes.
Light is propagated until absorbed, with no imposed speed limit other than those due to opacity changes (e.g. ionization fronts). 
{\sc trevr2} searches outward from receiving gas in discrete directions set by the {\sc healpix} algorithm (unlike its 
slower predecessor {\sc trevr}), 
accumulating optical depth and adding the flux due to sources combined into progressively larger tree
cells with distance.  We demonstrate $N_\textrm{active}\log_2 N$ execution time {\it with absorption} and many sources. 
This allows multi-band RT costs comparable to tree-based gravity and hydrodynamics, and the usual speed-up when active particles evolve on individual timesteps.  
Sources embedded in non-homogeneous absorbing material introduce systematic errors.  We introduce transmission averaging instead of absorption averaging which dramatically reduces these systematic effects.  We outline other ways to address systematics including an explicit complex source model.  We demonstrate the overall performance of the method via a set of astrophysical test problems.
}
\end{abstract}

\begin{keywords}
radiative transfer -- methods: numerical
\end{keywords}



\section{Introduction}\label{sec:intro}

Light is central to astronomy and astrophysics.  However, radiative transfer (RT) is usually treated very approximately in numerical simulations of astrophysical systems.  This is due to light's high intrinsic propagation speed and multi-dimensional nature.  RT can dramatically increase the computational time, forcing simulators to dramatically lower the resolution or scope relative to non-RT simulations which might otherwise define the state of the art in fields such as galaxy formation and making comparisons to prior work more difficult.  

Radiation can add momentum and complex dynamical interactions with matter (Radiation Hydrodynamics, RHD) but these effects mostly apply extremely close to sources (e.g. Stellar Winds) and are typically unresolved in galaxy-scale simulations.  More ubiquitous effects include ionization (chemical changes) and heating.  For example, UV and X-rays are important over a huge range of galactic and cosmological scales \citep{tielens05,wolfireEt03,puchwein2019}.
Radiation intensities vary by many orders of magnitude and yet it is common to use simple heating prescriptions (see however, \citealt{benincasa2020}).  
Thus, there is considerable value in being able to cheaply approximate the mean intensity and ionization rates due to distributed sources.  While angular precision (e.g. sharp shadows) is a key consideration for geometric optics, in practice scattering due to dust (where scattering opacity is similar to absorption) and recombinations that produce new ionizing photons diffuse the field direction and fill shadows \citep{hasegawa2010}.  This effectively makes all gas a source. 
The focus of this work is to consider fast, approximate RT methods that can handle large numbers of sources comparable to the number of elements, primarily for the purposes of gas chemistry and heating.

\subsection{Radiative Transfer Methods}

We begin by considering the full RT equation \citep[e.g.][]{mihalasMihalas84},
\begin{eqnarray} \label{eqn:classicrt}
\left[ \frac{1}{c} \frac{\partial}{\partial t} + \mathbf{n \cdot \nabla}
 \right] I\left(\mathbf{x}, \mathbf{n}, t, \nu\right) =
\epsilon\left(\mathbf{x}, t, \nu\right) \nonumber \\
+ \frac{1}{4\,\pi} \alpha_\textrm{s}\ J\left(\mathbf{x}, t, \nu\right) - (\alpha_\textrm{abs}+\alpha_\textrm{s})\  
I\left(\mathbf{x}, \mathbf{n}, t, \nu\right).
\end{eqnarray}
Here, $I$ is the intensity which depends on position $\mathbf{x}$, unit direction of light propagation $\mathbf{n}$,
time $t$ and frequency $\nu$. The quantities $\epsilon$, $\alpha_\textrm{abs}$ and $\alpha_\textrm{s}$ are the emissivity,
absorption and (isotropic) scattering coefficients respectively. These are commonly assumed to depend on position, time and frequency indirectly as properties of the material present. The mean intensity, $J=\int_\Omega I\, d\Omega$ integrates over angle and is the essential input for ionization and heating rates and also acts as source term for scattered light.  We have neglected redshifting and non-monochromatic scattering which may shift frequencies.
Apart from being a seven dimensional problem, RT involves the highest 
possible characteristic speed, $c$, the speed of light. Also, unlike a similar long range
interaction such as gravity, RT depends on the properties of the 
intervening material via the absorption and scattering terms.

Following the original {\sc trevr} paper \citep{grond2019}, we differentiate between {\it evolutionary} and {\it instantaneous} methods.  {\it Evolutionary} methods {store} the radiation field {on local resolution elements.  It may be treated} as a fluid expanded in moments {or another discrete representation, such as packets. The field is evolved by transporting energy from element to element at the speed of light}.  The proto-typical, zero-eth order method is flux-limited diffusion \citep{levermorePomraning81}.  Modern approaches use higher moments, e.g. {\sc otvet} \citep{gnedinAbel01}.  A popular approach is the first order M1-closure \citep{gonzalez2007} used adaptively in {\sc ramses-rt} \citep{rosdahlEt13}.  
{These approaches have limited angular resolution which creates problems with accurate shadows, consistent directions and crossing beams \citep{davis2012}.}
The most accurate schemes, such as the short characteristics-based closure in {\sc athena} \citep{davis2012}, {can employ many directions} per element{\bf, which can resolve the issues listed above via increased computation}.  The severe speed of light time step restrictions can be ameliorated somewhat with a reduced speed of light (e.g. \citealp{katz2022}){\bf, but will still require many radiation steps per hydrodynamic step in general}.  Assuming a system with $N$ elements (cells or particles), such methods have similar scaling to hydrodynamics as $\bigO{N}$ and are well suited to RHD.   Radiation can also be transported as packets as in {\sc traphic} \citep{pawlikSchaye08}, {\sc sphray} \citep{altayEt08} and {\sc SimpleX2} \citep{paardekooperEt10}, which {are somewhat in the mode of a Monte Carlo methods and thus } noisier. {\it Evolutionary} methods typically require evolving the entire radiation field regularly to keep it up to date.

{\it Instantaneous} methods typically {\it ray trace} the light until it is absorbed.  {We note that ray tracing is used within some evolutionary methods to help map out the radiation field locally, such as to construct the radiation pressure tensor and thereby accurately transport the field.  We distinguish {\it instantaneous} methods as those where the entire calculation is the ray trace, over whatever distances are required, and no additional transport step is used.}  In principle, this limits the frequency that radiation has to be computed to the timescales of the sources and ionization fronts, which are typically long compared to light crossing times except extremely close to sources.  These methods may require iterations to converge if opacities are changing.  Such methods are {of restricted use} for RHD, as the {speed of light is effectively infinite and light waves don't propagate correctly.  However they are well suited to cases where ionization and heating set the characteristic time-scales.} 
Initial work focused on forward ray tracing from each of the $N_\textrm{src}$ sources, e.g. {\sc C$^2$-Ray}, \citep{mellemaEt06a}.  A characteristic distance is $N^{1/3}$, in terms of the number of elements traversed.  Casting rays from every source to every other element has an $\bigO{N_\textrm{src}\, N^{4/3}}$ cost.  
If each element casts rays once, with full angular resolution ($\bigO{N^{2/3}}$ rays), this reduces to $\bigO{N_{src} N}$. 
Costs can be reduced via hybrid characteristics (long \& short as in \citealt{rijkhorstEt06,klassenEt14}) or using angular adaptivity, such as via {\sc healpix} \citep{gorskiEt05} as used in {\sc moray} \citep{wiseAbel11}, {\sc fervent} \citep{baczynskiEt15} and \cite{kim2017} within {\sc athena}.  
This reduces the multiplier but not the overall $\bigO{N_{src} N}$ scaling as each of the $N$ elements must be intersected by at least one ray.

\subsection{Faster Ray Tracing}\label{sec:fastrt}

Forward ray tracing can {run at practical speeds} by limiting the number of sources.  {This is not possible if we have many bright sources (e.g. stars in a galaxy), or if most gas is a source, such as in the case of widespread recombination, line emission or scattering}.  
{We define active simulation elements as fluid cells or particles that are active on the current timestep.  The timestep can be set for each element individually by hydrodynamics or other processes such as chemistry so that only a subset need to be active at once}.
{However,} forward ray tracing necessitates tracing from all sources to ensure {the subset of} active elements {have an up-to-date radiation field}.  We would like to limit the {active} sources, ignoring those that are not currently relevant or merging distant sources.  This is most easily achieved via a reverse ray trace.  {For the purposes of this work, the key distinction of a {\it reverse ray trace} is that it can be performed per receiving (gas) element, rather than per source element (e.g. star).} 

Merging sources is equivalent to imposing a minimum beam size.   This is most useful if each beam can accumulate the radiation from separate sources that are within the same beam as viewed from the receiver.   Absorbers are treated once per beam as it reaches out toward the sources.  
The absorbers in this case may be large tree cells aggregating the properties of many elements.  This should scale as $\bigO{N_\textrm{receivers} \log_2 N}$.  

An early work with merged sources is {\sc start} \citep{hasegawa2010}, using a finite set of rays determined by an opening angle.  {\sc start} is an SPH code and begins RT at the scale of individual particles.  {\sc start} progressively combines rays from each source in parent tree-cells in a manner that is inherently serial in its current implementation.  The results display impressive scaling with the number of sources.  The test problems also demonstrate the diffusive nature of ionizing radiation and its ability to erase shadows when recombinations are included.  However, {\sc start} always computes the entire volume so that $N_\textrm{receivers} = N$ rather than $N_\textrm{active}$.  The {\sc argot} code \citep{okamoto2012} is presented as an adaptive mesh code implementation of {\sc start}.

Explicit reverse ray tracing codes include {\sc TreeCol} \citep{clarkEt12}, {\sc urchin} \citep{altayTheuns13}, {\sc treeray} \citep{wunsch2021, HaidEt18} and {\sc trevr} \citep{grond2019}.  These schemes typically use a tree to approximate distant sources, so that the effective number of sources is $\bigO{\log_2 N_\textrm{src}}$.  Reverse ray tracing can calculate RT to a subset of the receiving elements, allowing for adaptivity in time and space.   In principle, only sources that illuminate active regions must be fully counted and other sources can be treated approximately or ignored.  

Both the original {\sc trevr} scheme and that of the current work explicitly calculate radiation only for the $N_\textrm{active}$, active elements on the current hydrodynamics or ionization time step.  In galaxy formation, with a wide range of densities and temperatures, $N_\textrm{active}$ is typically a fraction of a percent of the total, on average (i.e. $N_\textrm{active} \ll N$).
The original {\sc trevr} scheme is thus $\bigO{N_\textrm{active} \log_2 N_\textrm{src}}$ when optically thin.  Absorption was also calculated using tree cells on a per source basis, giving $\bigO{N_\textrm{active} (\log_2 N)^2}$ overall performance.   
Larger tree cells can contain a mix of transparent and opaque subcells. Thus accurate shadows can require refinement. 
A novel feature of {\sc trevr} is a mechanism to adapt absorption where needed.  \cite{grond2019} showed that for large numbers of sources ($N_\textrm{src} \sim N$), the performance smoothly goes to $\bigO{N_\textrm{active} N^{1/3}}$ as the typical optical depth increases.

The overall cost of {\sc trevr}, as implemented in {\sc gasoline2} \citep{wadsleyEt17} and applied to galaxy-type problems with high opacities, is typically one to two orders of magnitude greater than the rest of the code.  This makes it relatively impractical for full-length simulations.   In the current work, we examine ways to limit this cost by progressively estimating absorption along a beam and applying it to all sources in that same beam.

\subsection{{\sc healpix}-based directions}\label{sec:healpix}

Given that source and absorber merging with an opening angle sets an effective angular resolution, it makes sense to limit the number of ray trace beams.  Evenly spaced directions (with equal solid angles) can be conveniently constructed using the {\sc healpix} scheme \citep{gorskiEt05}, such as in the forward ray-tracing codes {\sc moray} and {\sc fervent}; and the reverse-ray tracing codes {\sc urchin}, {\sc TreeCol} and {\sc treeray}.  Typically the basic set is $N_\textrm{ray}=12$ directions which can be multiplied by 4 raised to some power, based on a user specified parameter.

We focus on {\sc treeray} as a primary example of a reverse ray-tracing method with {\sc healpix}, because the algorithm is clearly described, the accuracy of the implementation is high and \cite{wunsch2021} include a varied set of well-defined tests which are easy to compare to.  
{\sc treeray} is implemented within the {\sc flash} adaptive mesh code \citep{FryxellEt2000}.  $N_\textrm{ray}=48$ is presented as a fiducial choice, which corresponds to beams with an angular size of $\sim$32 degrees, which is quite similar to the typical angle subtended by tree cells with a \cite{barnesHut86} opening angle of 0.5.  The opening angle should follow $(N_\textrm{ray}/12)^{-1/2}$ to maintain matched angular sizes.

{\sc treeray} samples the absorption in the beams with roughly 2 points per active cell using a weighted average of all cells in the beam.  The opening criterion ensures that active cell sizes increase in proportion to the distance from the receiver, making the overall number per ray scale roughly as $\log_2 N$.  The expected scaling thus should be $\bigO{N_\textrm{receivers} N_\textrm{ray} \log_2 N}$. 
The tests shown in \cite{wunsch2021} used relatively few sources, so that $N_\textrm{src} \ll N$ where $N = N_\textrm{src} + N_\textrm{absorbers}$.  As such, the demonstrated performance was largely independent of $N_\textrm{src}$.  

For the current work, we have developed a new scheme, {\sc trevr2}, which employs {\sc healpix} based directions to make absorption more efficient.  In section~\ref{methods},  we describe {\sc trevr2} as implemented within the {\sc gasoline2} Smoothed Particle Hydrodynamics (SPH) code.  We highlight key differences between {\sc trevr}, {\sc treeray} and {\sc trevr2}, including improved treatments for  optically thick elements (\ref{sec:optthick}), behaviour near sources (\ref{sec:intint}) and finding the effective opacity of merged cells (\ref{cellavg}). 
In section~\ref{sec:tests}, we evaluate the behaviour of {\sc trevr2} using test problems and demonstrate the expected performance on a realistic problem.  We include new test problems that highlight systematic issues related to the fixed directions approach.  We also present different approaches to overcoming the systematic issues.  Finally, we discuss lessons learned and future prospects in section~\ref{sec:discussion}, before we conclude in section~\ref{sec:con}.

\begin{figure}
    \centering
	\includegraphics[width=\columnwidth]{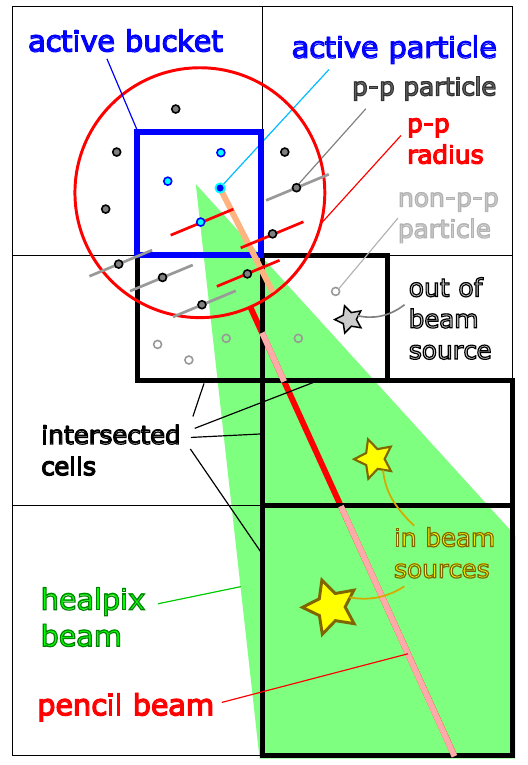}
    \caption{Schematic illustration of the {\sc trevr}2 method.  Particles are organized in buckets that are combined to form tree cells.  Absorption due to nearby particles is handled directly (p-p).  {The cross-section of each particle is indicated with a line perpendicular to the ray direction.}
    Absorption beyond the p-p radius is handled via intersecting cells along the centre of the healpix beam and using an effective absorption for each cell.  Only cell sources with a centre of luminosity in the healpix beam contribute along that ray. }
    \label{fig:method}
\end{figure}

\section{The {\sc trevr2} Algorithm}\label{methods} 

{\sc trevr2} employs radiative transfer along rays as illustrated in figure~\ref{fig:method}.  The smallest cell in the tree is a bucket containing a few gas particles (our receiving elements),  which are calculated together. The incoming rays are traced from the receiving particle out toward the sources.  Close to the receiving particle, we use particle based opacities and then switch to cells outside the bucket at the particle-particle (p-p) radius.

Sources within the p-p radius are treated individually (for example star particles). For heating and chemistry, we must apply an averaged local intensity that takes into account the overall size of the absorbing element as it affects optical depth (section~\ref{sec:optthick}) and the fact that rays will emit into a large solid angle when the source gets close to the absorber (section~\ref{sec:intint}).  

More distant sources are aggregated into a combined cell source as the tree is built, as described in section~\ref{sec:tree}.  
In the original {\sc trevr} a ray would be cast directly to this source.   In {\sc trevr}2, aggregated sources are assigned to one {\sc healpix} beam and treated in sequence, leading to a considerable speed-up, described previously in section~\ref{sec:healpix}.

As shown in figure~\ref{fig:method}, nearby absorbing particles are treated directly with a particle-particle ray trace, described in section~\ref{pp}.  
Beyond the p-p radius, we switch to a common {\sc healpix} beam for absorption for all particles in the bucket, as described in section~\ref{sec:pcell}.  
A key new aspect compared to the original {\sc trevr} is matching the cell angular scale to the {\sc healpix} beam.   
A key improvement for {\sc trevr}2 is how we combine the opacities in each cell, described in section~\ref{cellavg}.

We note that many of the improvements can be used in any ray-tracing code without requiring the use of {\sc healpix} directions.

\subsection{Optically thick elements}\label{sec:optthick}

In astrophysical simulations, optical depths can be very high on the scale of a single gas element.  For example, ionizing radiation at 13.6 eV has an optical depth of $\sim 20$ across a distance of $1$ pc for a density of $1$ Hydrogen atom per cubic centimetre.  This corresponds to a mass resolution of 0.1 M$_{\odot}$ per element, which is beyond that of current galaxy simulations.   Thus it will be common for the intensity to vary dramatically on the scale of one element.

A gas element with a mass, $m$, and density, $\rho$, has an effective volume of $h^3 = \frac{m}{\rho}$. $h$ is the inter-particle spacing for particle elements and the side-length for cubic cells.  We can approximate this as a simple shape with cross-section $h^2$ and length $h$.  Plane-parallel radiation with intensity $I(0)$, arriving at one face is attenuated according to 
\begin{eqnarray}
    I(x) = I(0) e^{-\tau(x)},
\end{eqnarray}
where $\tau'(x)=\int_0^x \alpha(x') dx'$ is the optical depth at point x inside the element with a maximum value across the {element} of $\tau=\tau'(h) = \langle\alpha\rangle h$.   We assume that heating and reaction rates are proportional to $I(x) \alpha(x)$ at each point and find the average,
\begin{eqnarray}
    < I\ \alpha >_\textrm{element} = \frac{1}{h} \int_0^h I(x)\,
    \alpha(x)\, dx = \nonumber \\
    \frac{1}{h} \int_0^{\tau} I(0) e^{-\tau'} d\tau' = \frac{1}{h} I(0)\ (1-e^{-\tau}). \\
    \textrm{Alternately\ } <I>_\textrm{element} = I(0) \frac{(1-e^{-\tau})}{\tau},\label{taucorr}
\end{eqnarray}
Equation~\ref{taucorr} gives the average intensity experienced by the gas in the case of a constant $\alpha$ in the element.  In the limit of small $\tau$ this approaches {$I(0) (1-\tau/2)$, which is the outcome from using} the midpoint value, $I(0) e^{-\tau/2}$. However, at large optical depths the {correct} average intensity scales as $1/\tau$ and is much larger than the midpoint estimate.  

If we integrate {the rate} over the element volume, $\int_\textrm{element} I\, \alpha\, d^3x \rightarrow  h^2 I(0)$ as $\tau\rightarrow \infty$. {Here the volume element $d^3x = h^2 dx$ and otherwise we are using the expressions from eqn~\ref{taucorr}.} In this limit, all the photons entering the element {through the face with area $h^2$} are absorbed {as required}.

A related point is that the RT should provide the intensity striking the element surface, $I(0)$, without self-absorption by that element.  This is critical or the element may prevent its own ionization by attenuating the radiation.

\begin{figure}
    \centering
	\includegraphics[width=\columnwidth]{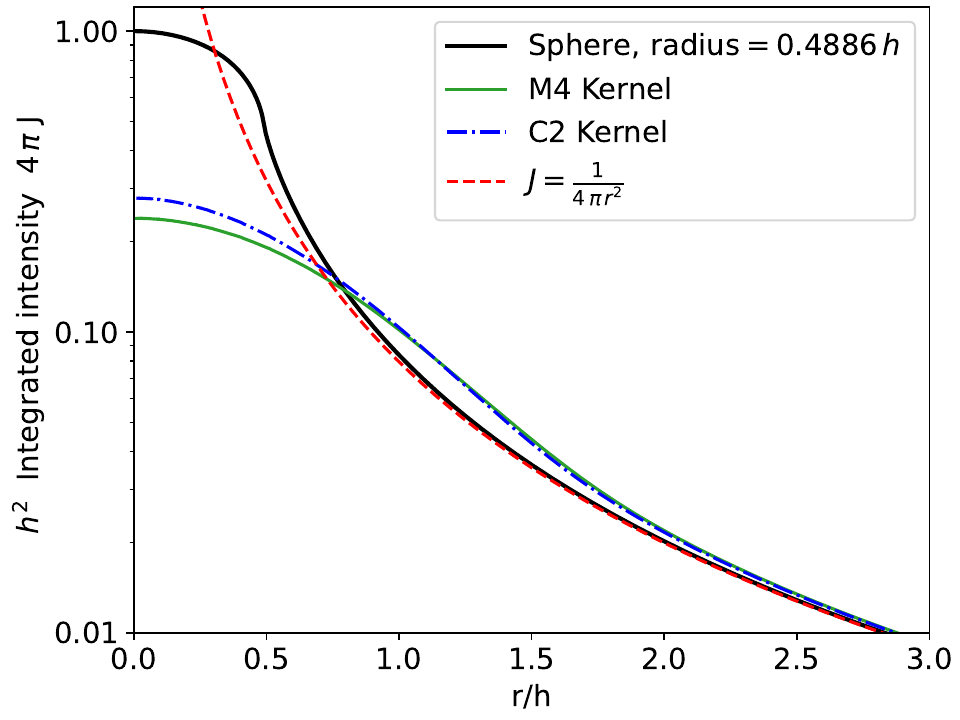}
    \caption{Element averaged integrated intensity versus distance from a unit luminosity, point source.   Integrated intensity exceeds $\frac{1}{4\,\pi\,r^2}$ as angular size gets large ($r \sim 1-2\,h$), then smoothly goes to a maximum at zero separation.  The preferred curve is the exact solution for an homogeneous absorbing sphere of radius 0.4866 times the particle spacing, $h$ {(eqn~\ref{intint})}.  {The remaining curves indicate the numerically integrated exact results for the average intensity weighted in space according to two popular choices of SPH kernel (M4 and C2).}}
    \label{radsoft}
\end{figure}

\subsection{Integrated intensity versus distance:  Finite size elements}\label{sec:intint}

The element averaged integrated intensity is the input for chemistry and heating rates.  {\sc trevr2} considers sources to be points.  At large distances, the angular distribution of rays from a source approaches a delta function.  In this case flux and intensity are interchangeable and follow the standard inverse square relation. When a source is nearby it produces intensity in a range of angles.  While the flux falls to zero for a source at the element centre, the integrated intensity increases smoothly to a maximum as the source moves inward.  This is not numerical softening, but the correct outcome for a finite size element.  We apply the following functional form for the integrated intensity, {$\int I\ d\Omega = 4 \pi J$}, averaged within an element without absorption, 
\begin{eqnarray}
    < 4 \pi J >_{\textrm{element}}(r,h) = L\
          \frac{1}{2\,h^2} \left( 1 + \frac{x^2 -1}{4 x} \ln{\frac{(x-1)^2}{(x+1)^2}} \right),  \nonumber \\
    \textrm{where \ } x = \frac{r}{0.4886\, h}.\label{intint}
\end{eqnarray}

This is the black curve shown in figure~\ref{radsoft}.  {Equation~\ref{intint} is the volume average of the mean intensity at all points within a sphere of radius $(\frac{3}{4 \pi})^{1/2}\,h \sim 0.4886\,h$ due to a point source of luminosity, $L$, offset from the sphere centre by a distance, $r$}.  
{This is not intended as a precise representation of the density distribution in the fluid element.  The radius of the sphere, was selected to give the correct peak} average intensity of $\frac{1}{h^2}$ at $r=0$.  {It can be interpreted as physically motivated softening function, suitable for radiation from a point source}.  For an optically thick element, after the intensity correction from eqn.~\ref{taucorr} is applied, the element will thus absorb all the energy, $L$, from an embedded source at $r=0$. The integrated intensity closely approaches the expected inverse square law for $r \gtrsim 2\, h$.

Note that SPH particles can be interpreted as loosely extending out to $2-3$ particle spacings in all directions (e.g. the standard M4 and Wendlend C2 kernels also shown in Figure~\ref{radsoft}, with support of $2 h$).  However, in this case,  particles may {be} partially or completely shielded by other particles that are {mostly behind} them and erroneously halt ionization fronts.   Therefore, we treat particles as compact, non-overlapping absorbing volumes for RT (see also \citealp{altayTheuns13}).

\subsection{Radiation Tree}\label{sec:tree}

Radiative transfer is computed using a binary tree, where each division is half-way along the longest axis.  In general, this means each tree node is cubic or at most has an axis ratio of 2:1.  The tree is built every radiation step and represents a fixed amount of work that scales as $N \log_2 N$.   The smallest cells are buckets which contain a user specified maximum number of particles (default 12) or fewer which is similar to the way {\sc gasoline}2 builds trees for other purposes (e.g. SPH and gravity).  Thus the spatial resolution is comparable to half an SPH smoothing distance or a typical particle spacing.  Once built, the work to combine opacities and luminosities in the tree scales as $N$ and is thus relatively cheap.  Combining methods have important consequences, as discussed in detail in section~\ref{cellavg}.  The tree build is evenly load balanced in parallel.  The cost of the tree build is typically roughly a percent of the total cost of radiative transfer.  However, this limits the potential gains from adaptive time steps to a factor of 100 or so because eventually the tree build is more expensive than the rest of RT on the shortest timesteps.

\subsection{Particle-Particle Radiative Transfer}\label{pp}

{\sc gasoline2} is an SPH code, so the elements are particles.  Sources are traced directly to particles on the scale of a few particle spacings, $h_i = (m_i/\rho_i)^{1/3}$, where, respectively, $m_i$ is the mass and $\rho_i$ is the density for particle $i$.  
A sphere that extends slightly beyond the containing bucket for each particle denotes the cut-off distance, $r_\textrm{p-p}$ from the bucket centre.  This radius is also increased until no p-p particle is larger than the {\sc healpix} beam at this distance.  
As shown in figure~\ref{fig:method}, this sphere will partially intersect some nearby cells and all particles within this radius are treated directly using p-p.  Remaining particles are instead treated via their contribution to the containing cell, discussed in section~\ref{sec:pcell}.

Particles, $j$, are considered to be entirely in front if they are closer to a source than the receiver $i$, ensuring a unique ordering \citep{grond2019}. Such particles are indicated with a line in the figure, which is colored red if they also intersect the ray to the source.  The intersected mass column, $\Sigma_j(b)$, is reduced linearly with the transverse distance, $b$, divided by $h_j$ and normalized so that an area integral from $0$ to $h_j$ accounts for the entire particle mass, $m_j$.  The received intensity is reduced by a factor $e^{-\kappa_j\, \Sigma_j(b)}$ for every particle $j$ in front of the receiver with opacity, $\kappa_j$.  This is similar to pure particle RT schemes (e.g. {\sc sphray}).   
Beyond $r_\textrm{p-p}$ absorption is estimated from cell properties. 

\subsection{Particle-Cell Radiative Transfer}\label{sec:pcell}

Rays are traced from receiving buckets if that bucket has one or more active gas particles.  Beyond the p-p radius, a single pencil beam is used, traced from the bucket centre and starting from the p-p radius.  {\sc trevr2} uses directions defined by the {\sc healpix} tessellation of the sphere \citep{gorskiEt05}.  The total number is a user defined parameter, with the default value $N_\textrm{ray}=48$.  It is most efficient to walk the entire tree for each bucket and assign cells to one or more {\sc healpix} directions in one pass, which is similar to the {\sc treeray} approach.  

\subsubsection{Cell Ray Trace}\label{sec:cellraytrace}

Each of the $N_\textrm{ray}$ directions has its own priority queue (ranked by distance from the receiver).  Cell sources are tested using various opening criteria (discussed in detail in sec.~\ref{sec:cellopenref}).  If they pass, they are added to the queues according to the distance from the bucket centre to the centre of luminosity of the cell projected onto the {\sc healpix} direction.   Each source is added to exactly one {\sc healpix} ray's queue.  Adding an entry to a priority queue is $\bigO{1}$, so the tree walk is $\bigO{N_\textrm{active}\, \log_2 N}$ overall without refinement.

Cells with non-zero opacity are added to one or more queues according to the closest point of intersection along the ray from the bucket centre in the appropriate {\sc healpix} direction as shown in figure~\ref{fig:method}.  This ensures that all relevant absorbers are ranked ahead of the sources in the queue.  Only the part of the rays beyond $r_\textrm{p-p}$ are considered.  Absorption within $r_\textrm{p-p}$ is added using the particle-particle approach from section~\ref{pp}. 

To trace each {\sc healpix} ray and calculate received intensities, we start with the queue entry closest to the receiving bucket.  Extracting entries from the queue is $\bigO{\log_2 N}$, thus overall we achieve the target $\bigO{N_\textrm{active}\,N_\textrm{ray} \log_2 N}$ scaling.  Absorbing cells are popped from the queue and added to the total optical depth of the ray.  If a source is found, only the ray up to the distance to the source is considered at that time.  
The projected distance along the {\sc healpix} direction is shorter than the true receiving particle to source distance.  This is corrected by dividing the optical depth by this ratio on a particle-by-particle basis within each bucket (which always makes the optical depth larger, as expected).  That source's contribution to the integrated intensity experienced at the receiving particle is then computed using equation~\ref{intint}.  If desired, the code could calculate fluxes (for radiation momentum) or take into account the total absorbing column (e.g. for complex transfer such as Lyman-Werner bands) at this time.

\subsubsection{Cell Opening Criteria and Refinement}\label{sec:cellopenref}

{\sc trevr2} employs a cell opening criterion $B/d > \theta_\textrm{rad}$ similar to gravity in Gasoline \citep{wadsleyEt04}, based on the distance, $d$, from the receiving bucket centre to the closest cell edge, not the centre, and the largest distance $B$, from the centre of luminosity to the cell edge.  This ensures a cell must always open itself for any $\theta_\textrm{rad}$.  For the default $\theta_\textrm{rad} = 0.75$,  a cubic cell with a centred source subtends a half-angle of 16.8-23 degrees, similar to the average half-angle for 48 {\sc healpix} cells (16.5 degrees).  It thus behaves similarly to the {\sc treeray} Barnes Hut criterion of $\theta_\textrm{lim} = 0.5$.  This opening criterion leads to $\bigO{\log_2 N}$ cells being considered.  It directly constrains errors associated with the quadrupole moment of multiple sources, very similar to a tree gravity calculation.

As noted by \citet{wunsch2021}, different criteria can be applied for cells sources and absorption.  
A particular innovation for {\sc trevr} \citep{grond2019} was that the accuracy of the absorption could be refined by opening cells so as to keep the variation in optical depth with the cell less than a user specified parameter, $\tau_\text{refine}$.  This criterion was demonstrated to ensure sharp shadows in {\sc trevr}.  However, it makes the code much slower for highly variable optical depth, to the extent of forcing all cells to be opened if $\tau_\text{refine} \ll 1$.  Refinement is off by default in {\sc trevr2} ($\tau_\textrm{refine} = \infty$).

{\sc trevr} and {\sc trevr2} both use pencil beams that intersect cells.  In the original {\sc trevr} approach, beam angular sizes were not fixed in advance but instead set by the angular size of the cells that were used.  Cells could be refined as required for a specific problem without any systematic issues because the rays are always directed at the source (see \citealp{grond2019}, section 3.2.2 particularly).  For {\sc trevr}2 with the default cell opening parameter, the selected cells have an angular scale similar to the {\sc healpix} beam so that these pencil beams sample the absorption on the beam scale.  {\sc treeray} takes the extra step of averaging the cell properties onto sample points in the beam which ensures that the averages are always on at least the beam scale.  Thus {\sc trevr2} and {\sc treeray} are similar as long as the cell sizes are similar to the local beam width.   However, if cells were to be refined so as to be smaller than the {\sc healpix} beam, this can lead to a mismatch in the effective beam sizes.  In particular, the direction to the sources and the ray through the absorbers may differ by more than the effective angular resolution, with consqeuences examined in sections~\ref{sec:tests} and ~\ref{sec:discussion}.

\begin{figure}
  \centering
  \begin{minipage}[b]{1.0\linewidth}
    \centering
    \includegraphics[width=\columnwidth]{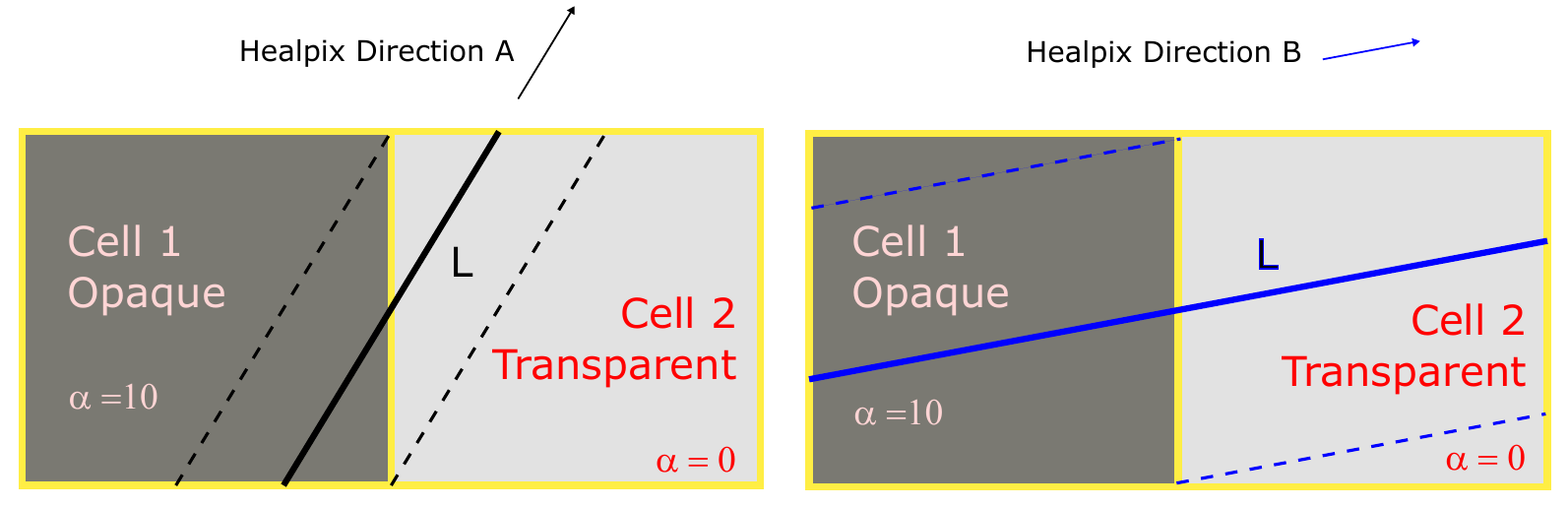}
    \qquad   
	\resizebox{\columnwidth}{!}{
 \begin{tabular}{llccl} 
		\hline
		  Method & Direction & $\alpha_\textrm{combined}$ & $\tau_\textrm{eff}$ & Transmitted \\
		\hline
		  Absorption average & A & 5 & 5.7 & 0.0032 \\
          \  & B & 5 & 10.4 & 0.000030\\
          \ & Thinnest (vertical) & 5 & 5 & 0.0067 \\
        \  & Thickest (diagonal) & 5 & 11.2 & 0.000014\\
        
		 Transmission average & A & 0.9 & 1.03 & 0.357  \\
		  \ \ (eqn.~\ref{transmissionavg}) & B & 1.4 & 2.91 & 0.05447 \\
        \  & Thinnest (vertical)  & 0.693 & 0.693 & 0.5 \\
        \  & Thickest (horizontal) & 5 & 10 & 0.000045  \\
        		\hline
	\end{tabular}
 }
  \end{minipage}
  \caption{Illustration of the transmission of rays through two cells, where cell 1 is optically thick ($\alpha_1=10$) and cell 2 is transparent.  The cells have side length 1.  Cell 1  blocks over 99.99 \% of the light passing through it.  However,  if the two cells exist within a beam, depending on the angle, up to half the rays will be transmitted through the transparent cell so that typically only 65-95 \% of the light would be blocked as shown in the table, giving a high average transmission overall.  Linear absorption averaging produces results similar to the worst case direction all the time.  }
  \label{fig:alphacomb}
\end{figure}

\subsection{Absorption: Transmission Average}\label{cellavg}

The absorption coefficient for an SPH particle is $\alpha = \kappa/\rho$ where $\kappa$ is the opacity and $\rho$ is the density. When averaging SPH particle properties into buckets (the smallest cells), it is important to weight particles according to their effective volume $V_i=m_i/\rho_i$, so that, 
\begin{eqnarray} 
\alpha_\textrm{cell} =\frac{\sum_i \alpha_i V_i}{\sum_i V_i} =  \frac{\sum_i \kappa_i m_i}{\sum_i m_i/\rho_i}.
\end{eqnarray}
This avoids discreteness noise on the bucket scale ($\approx$ SPH smoothing scale).

Above the bucket scale, absorption coefficients are combined into progressively larger cells.
A key choice for {\sc treeray} is to use weighted, {\it linearly averaged absorption} coefficients at all stages, beginning with individual grid cells being combined into tree nodes and also for the ray sample points (with various weighting functions as described in \citealt{wunsch2021}).  We initially tried  linearly averaged absorption,  but we found this has systematic effects and now use a {\it transmission average}.  

When the medium is clumpy, with mixed high and low opacities on scales smaller than the beam, the linearly averaged absorption approach will tend to over-estimate the typical opacity in the beam.  
An example case is that of a large tree node that is half-blocked by completely opaque material (e.g. $\tau = 1000$).   Half the radiation should pass through (an effective optical depth of $\tau = -\ln{0.5}$) but a linearly averaged absorption will result in $\tau = 500$ and no radiation will pass.  We illustrate how even moderate optical depths are affected in figure~\ref{fig:alphacomb}. 

In the original {\sc trevr} scheme \citep{grond2019}, the refinement scheme can adaptively use smaller cells until the issue is minimized, at the cost of more computation, which is equivalent to adaptively shrinking the beam angular size.  Thus the original {\sc trevr} could do well with linearly averaged absorption.  For the {\sc healpix}-based schemes discussed here, the beam angular sizes are fixed and on-the-fly adaptivity is less effective.  We compare with and without refinement in our tests in section~\ref{sec:tests}.

An approach more in keeping with how radiation is attenuated is to use a {\it transmission average} in each direction.   Transmission is the fraction of the original ray intensity that successfully traverses the cell (effectively $e^{-\tau}$ for that ray). 
As the tree is constructed, cells are combined pairwise.  We can exactly calculate the volume weighted fraction of rays that pass through each cell and which ones pass through both.  We can then estimate the average transmission of radiation,
\begin{eqnarray}
    \exp(-d\ \alpha_\textrm{cell, ray}) = \frac{1}{V} \Big( V_1 \exp(-d\  \alpha_1) + V_2 \exp(-d\ \alpha_2)  \nonumber \\  + V_{12} (\exp(-\frac{1}{2} d\ \alpha_1)+ \exp(-\frac{1}{2} d\ \alpha_2) ) \Big), \label{transmissionavg}
\end{eqnarray}
where the subscripts 1 and 2 refer to properties of the two child cells of the final cell.  $V_1$, $V_2$, $V_{12}$ and $V$ refer to the volume of rays passing only through cell 1, only through cell 2, through both and the total volume of the combined cell respectively.  We can take the $\log$ of equation~\ref{transmissionavg} to find the transmission average: a non-linearly averaged effective absorption $\alpha_\textrm{cell, ray}$ for that direction through the combined cell.  Example quantitative outcomes for transmission averaging are shown in figure~\ref{fig:alphacomb}. 

Equation~\ref{transmissionavg} uses the volume-averaged transmission, which is equivalent to an average over rays when the lengths are the same.  This length is the intersection length, except for rays through the cell corners.  We use the transmission of a ray passing through the centre of the face joining the two cells with half its length in each cell to represent all rays passing through both.  The distance, $d$, is the total length of this intersecting ray.  
This length and the volume weights change with each {\sc healpix} direction.  Symmetries for rays in the forward and backward direction lead to $N_\textrm{ray}/2$ stored $\alpha_\textrm{ray}$ values per cell. 

By expanding equation~\ref{transmissionavg} to first order in $\tau$, we can see that for small $\tau$ or small variations $|\tau_1-\tau_2| \ll 1$,  it gives the same result as a volume weighted, linearly averaged absorption.  There are caveats to this approach.  It assumes that the covering factor of opaque structures does not approach 100 \% and/or that the radiation tends to diffuse around obstacles on the scale of the beam width.  As noted in section~\ref{sec:intro} and demonstrated by \cite{hasegawa2010}, significant directional diffusion {can occur} in astrophysical scenarios of interest. Transmission averaging allows us to stay fast ($N \log_2 N$ scaling) relative to a refinement based approach.

\begin{figure}
    \centering
	\includegraphics[width=\columnwidth]{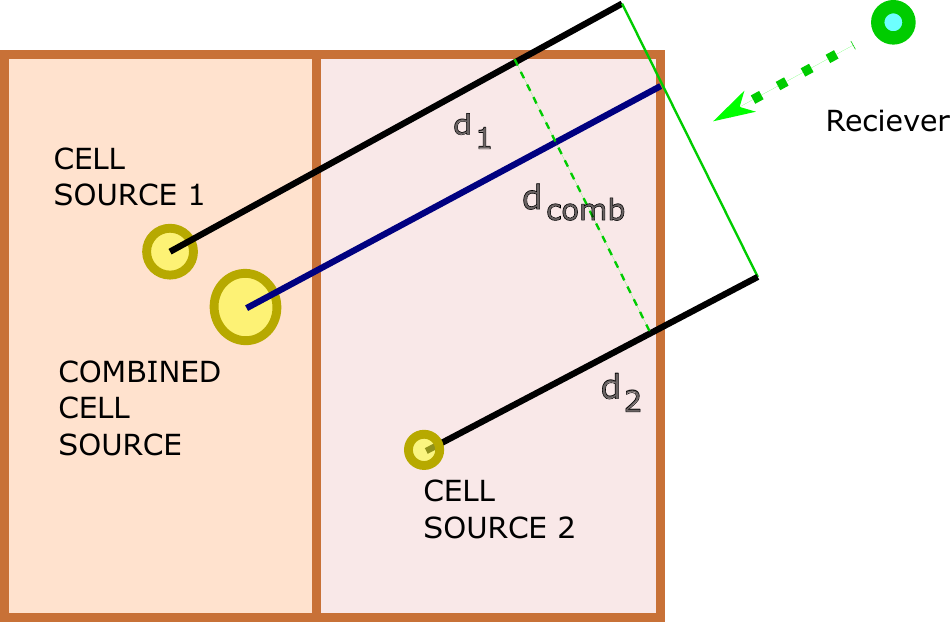}
    \caption{Illustration of combining sources. We add the two child cell luminosities and place the combined source at the luminosity weighed centre.  In the {\it Complex Source Model}, we must also combine the intrinsic absorption associated with the cells in each {\sc healpix} direction (depicted as rays toward the receiver).  The points of intersection of each ray with the cell edge are at different offsets when projected onto the {\sc healpix} beam and the lengths are adjusted so as to use the same projected location before being combined.}
    \label{fig:sourcecombine}
\end{figure}

\subsection{Complex Source Model}\label{sec:complexsourcemodel}

Once we began to test the above methods, we realized that compact structures including multiple sources and absorbers lead to complex angular variations in the radiation field far from the source that are hard to capture.  At these distances, cells would be combined and treated as a single source and the angular information is lost in standard {\sc trevr2}.  
In an attempt to address this without giving up on the fast $N \log_2 N$ scaling, we added an optional extension in the form of an angular model for the absorption around each cell source which we call the {\it complex source model}. 

Standard {\sc trevr2} already builds a ray-dependent transmission averaged absorption for rays passing through each combined cell.  The complex source model builds something similar for the absorption along rays aimed at the source within each cell and uses this instead of performing a full ray trace within that cell when calculating absorption in a {\sc healpix} beam as described in section~\ref{sec:cellraytrace}.  

With the complex source model, for each cell source, we store $N_\textrm{ray}$ absorption coefficients and ray lengths for each {\sc healpix} direction.  These apply to the rays connecting the combined source to the edge of the cell the source is in.   Rather than trace these rays repeatedly, when cells are combined, 
we extend the lengths using a new ray-segment from the edge of the child cell to a point near the edge of the larger, merged cell in each direction and recalculate the averaged absorption coefficient as illustrated in figure~\ref{fig:sourcecombine}.   
We use a transmission type average similar to equation~\ref{transmissionavg}, so that the final optical depth ($= \alpha_\textrm{comb,ray} d_\textrm{comb,ray}$) gives a combined luminosity with absorption that is the sum of the attenuated luminosity of the child sources along their own beams,
\begin{eqnarray}
L_\textrm{comb} e^{-\alpha_\textrm{comb,ray} d_\textrm{comb,ray} } = 
L_\textrm{1} e^{-\alpha_\textrm{1,ray} d_\textrm{1,ray}} +
L_\textrm{2} e^{-\alpha_\textrm{2,ray} d_\textrm{2,ray}}
\label{eqn:srccomb}
\end{eqnarray}

This approach is able to partially take into account differences in the absorption each child cell source sees prior to being combined.  Note that the geometric factor of $1/4 \pi r^2$ is applied later.

As depicted in the figure, we are intrinsically assuming that all rays are parallel to the {\sc healpix} beam direction. This approximation is consistent with the fact that this cell is roughly the size of the full beam when used by a receiver.   

The cell ray trace step calculates all absorber rays projected along the {\sc healpix} beam centre ray.  This distance is ultimately corrected to the true distance per receiver prior to flux calculation, as described in section~\ref{sec:cellraytrace}.  
To remain consistent with this distance projection, rays from the child sources must end at the same point as the combined ray when projected.  The three ray lengths are indicated with a $d$ in the figure.  There are multiple potential choices for the end point and we tested several approaches (dashed and solid green lines perpendicular to the rays in the figure).  The default uses the combined source ray to the box edge to define the projected stopping point (solid green line).  Thus the rays for the child sources must be clipped or extrapolated as necessary before being combined using equation~\ref{eqn:srccomb}.  This is why we store coefficients and lengths separately, rather than just the optical depth per ray.   

This adds a fixed amount of work to the existing work to combine a cell.  Even for this fairly simple model for a complex cell source, it is about 40 \% more wall clock time overall for the tree build.  A more elaborate model could be employed, such as performing a ray trace within sub-cells of the children, applying some opening criteria.  However, this would dramatically increase the tree build time.  With the current, simple approach, the tree build with the complex source model is somewhat more expensive, but still $\bigO{N}$.   During the reverse ray trace, we must trace from the source cell edge rather than the source location, which means slightly less ray tracing. 
Importantly, the $N_\textrm{active} \log_2 N$ scaling is preserved for the RT calculation. 

\begin{figure}
    \centering
	\includegraphics[width=\columnwidth]{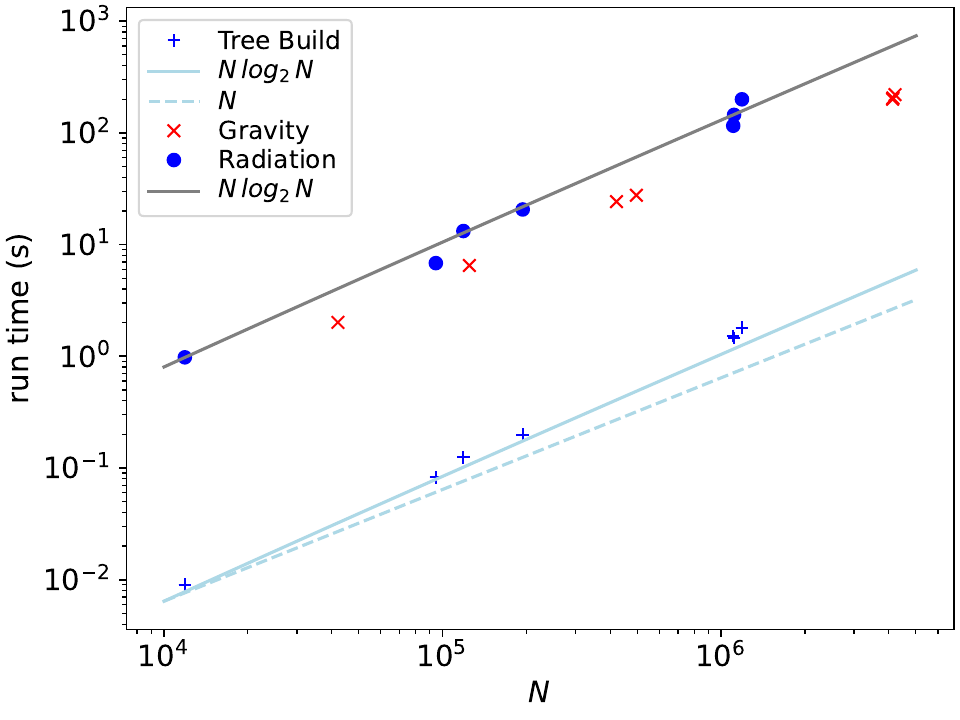}
    \caption{Basic scaling performance for {\sc trevr2} algorithm on a zoomed galaxy simulation {(one radiation band for a single force calculation), with run times in seconds}.  The largest $N$ values shown are from the original simulation.  For smaller $N$, the input simulation had particles removed.  The target $N \log_2 N$ RT scaling is clearly evident for RT (matching that of tree-based gravity and the tree-build, albeit with different pre-factors).  Note that the actual run time for RT is similar to gravity for each case.  The $N$ is larger for gravity only because dark matter also participates.  Thus the overall cost to add RT is comparable to running gravity.}
    \label{fig:scalingbasic}
\end{figure}

\begin{figure}
    \centering
	\includegraphics[width=\columnwidth]{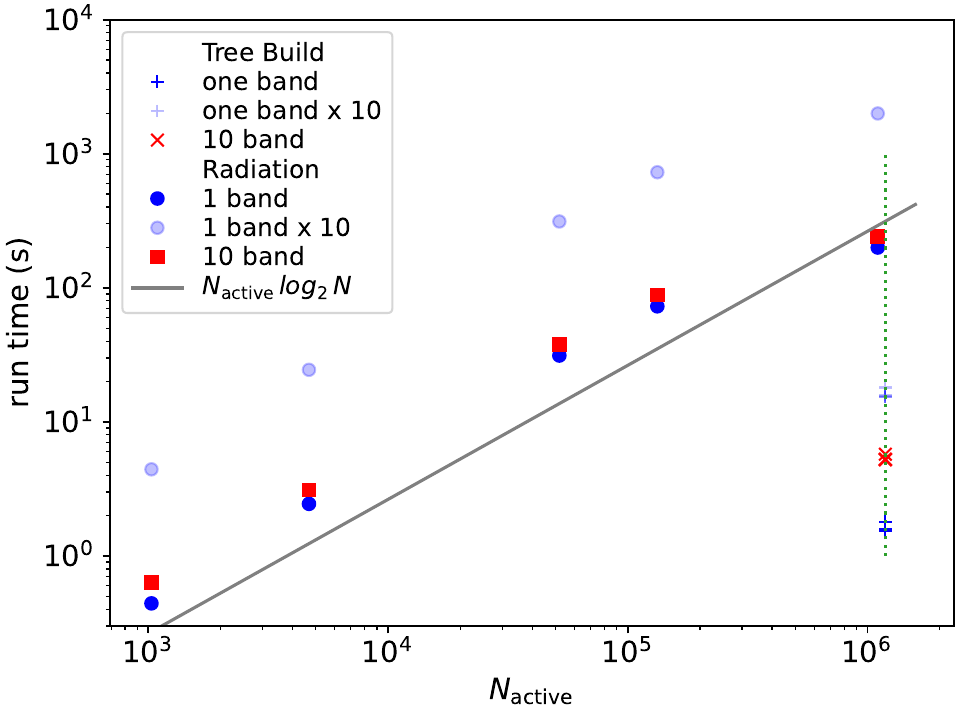}
    \caption{Scaling performance for {\sc trevr2} algorithm on a zoomed galaxy simulation with varying subsets of active gas particles updating their radiation field. The target $N_\textrm{active} \log_2 N$ RT scaling is clearly evident, particularly for smaller $N_\textrm{active}$. {The light blue circles and plusses are the single band multiplied by 10, representing doing radiation 10 times.} The red points demonstrate the small cost associated with adding radiation bands.  10 bands costs $1.3\times$ (red) rather than $10\times$ {(light blue circles)}.
    }
    \label{fig:scalingactive}
\end{figure}

\begin{figure}
    \centering
	\includegraphics[width=0.49\columnwidth]{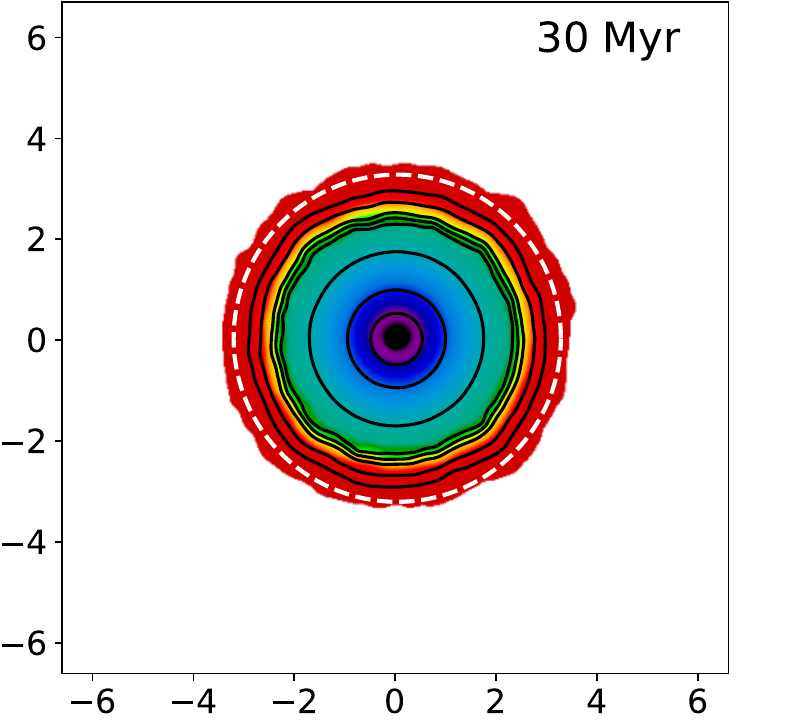}
     \includegraphics[width=0.49\columnwidth]{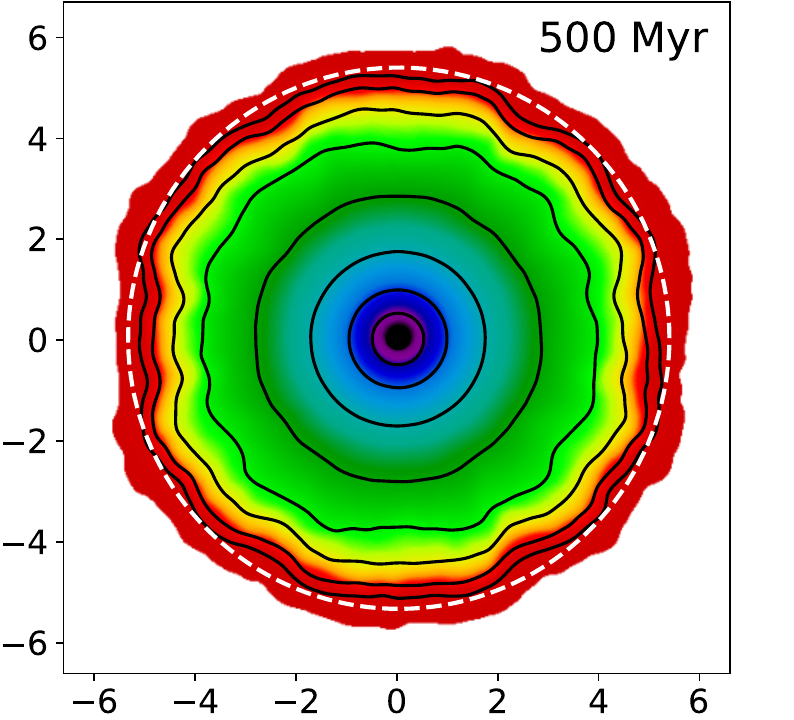}
    \caption{Isothermal Str\"{o}mgren test for {\sc trevr2} at 30 and 500 Myr 
    showing HI fractions with black contours at $10^{-4}$,  $10^{-3.5}$,  $10^{-3}$, 
    $10^{-2.5}$,  $10^{-2}$, $0.1$, $0.5$ and $0.9$.   The axis labels are in kpc. 
    The white dashed contour is the analytical radius estimate from \citet{grond2019}. The solution is slightly less spherical compared to the original {\sc trevr}.}
    \label{fig:stromiso}
\end{figure}

\begin{figure*}
    \centering
	\includegraphics[width=\textwidth]{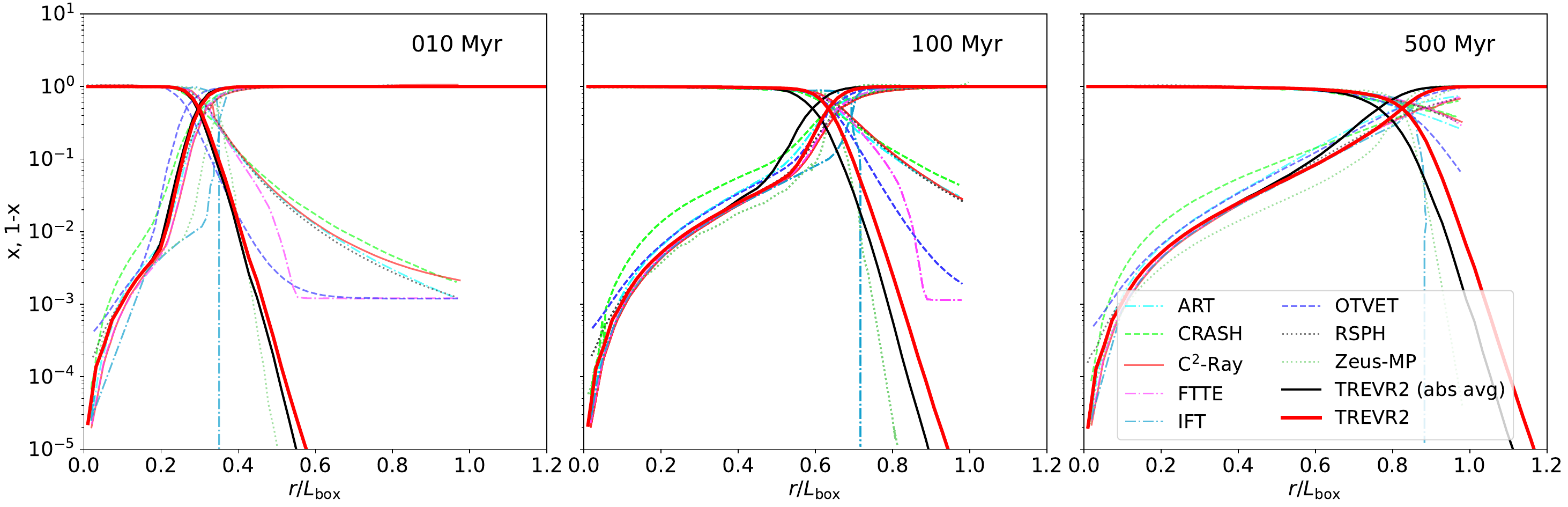}
    \caption{Str\"{o}mgren test with cooling.  {\sc trevr2} with transmission averaging (red curves) gives good results.  {\sc trevr2} with absorption averaging (black curves) underestimates the HII region size.  The other curves are taken from \citet{ilievEt06} and the labels are described there.}
    \label{fig:stromrad}
\end{figure*}

\begin{figure}
    \centering
	\includegraphics[width=0.8\columnwidth]{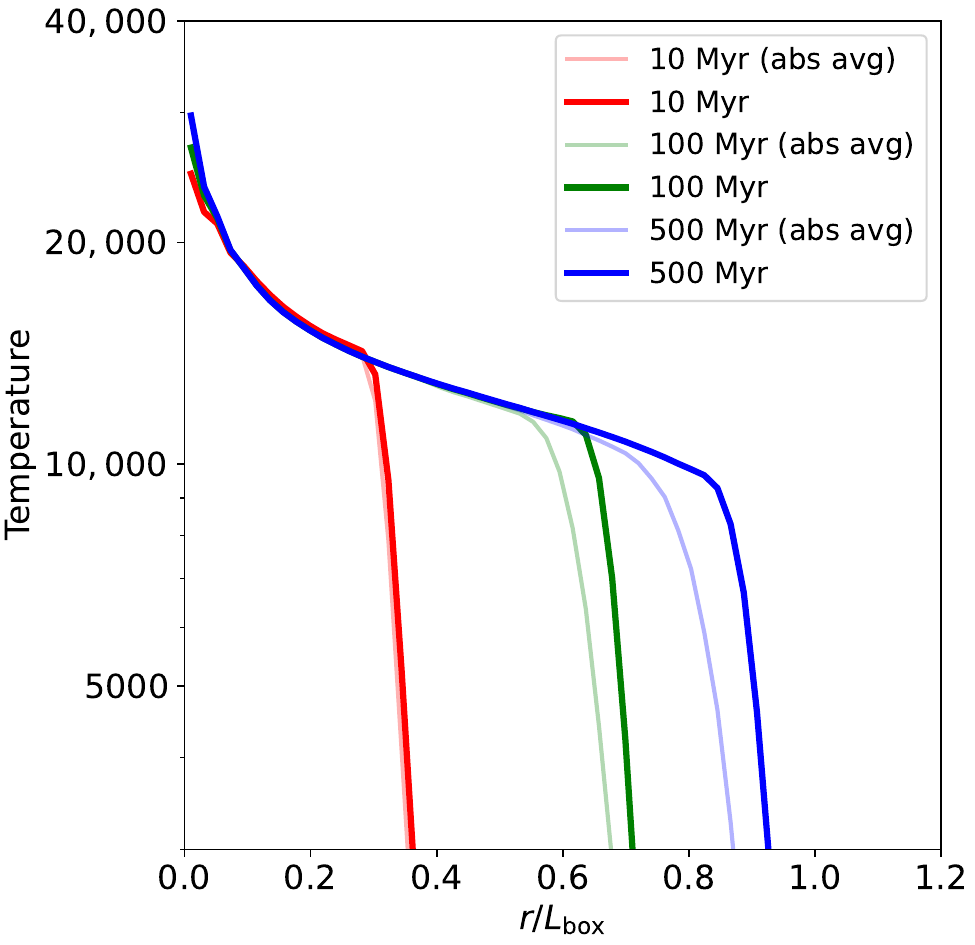}
    \caption{Temperature profiles for Str\"{o}mgren test with cooling.  The results are similar to {\sc trevr} (Grond et al. 2019) and other prior work.  The HII region sizes are small with absorption averaging (abs avg).}
    \label{fig:stromtemp}
\end{figure}

\section{Test Results}\label{sec:tests}

A primary goal of this work is the explore the utility of $N \log_2 N$ ray tracing methods.  Thus we begin by demonstrating the scaling and overall speed of the method.  We then show a standard test, the St\"omgren sphere, which is important but forgiving in some respects because it is spherical and has a low opacity interior, near the source.  {We note that the  {\sc trevr2} version used here does not include scattering or gas sources.  Thus we use case B recombination rather than explicitly creating new ionizing photons due to recombinations.}  We then replicate the two sources plus cloud test from \cite{wunsch2021}, and demonstrate systematics for non-symmetric tests.  We show additional tests, using several different approaches, to demonstrate the challenges  of finding a fast approximation that can handle complicated source geometries, i.e. retaining the $N \log_2 N$ scaling.  

\subsection{Galaxy simulation: Scaling}\label{sec:scaling}

The original {\sc trevr} paper used a perturbed particles set-up to test scaling.  Here we use a published zoomed galaxy simulation and measure the wallclock time directly.  The galaxy simulation is g1536 from the MUGS2 set of zoomed galaxies \citep{keller16} at redshift $z=0.26$.  The initial condition set up used a WMAP3 $\Lambda$CDM cosmology with $H_0=73\rm \,km\,s^{-1}\, Mpc^{-1}$, $\Omega_m=0.24$, $\Omega_\Lambda=0.76$, $\Omega_b=0.04$ and $\sigma_8=0.76$.  The simulation was originally run with {\sc gasoline2} \citep{wadsleyEt17}.  

In the high-resolution zoom region, the particle masses are $1.1\times10^6\,\mathrm{M}_\odot$ and $2.2\times10^5\,\mathrm{M}_\odot$ for dark matter and gas, respectively.  The impact of radiative transfer on this galaxy is studied in detail in a companion paper (Baumschlager et al., in prep).  Most important for scaling are the particle numbers and their distribution.  The original output had 3,021,673 dark matter, 1,104,175 gas and 83,544 star particles.  To examine how the scaling varies with the number of sources (star particles) and receivers (gas particles), we repeated the runs after deleting particles to reduce the numbers by factors of 10 and 100 as desired.   

In figure~\ref{fig:scalingbasic}, we show that we achieve the expected $N \log_2 N$ scaling for our implementation of {\sc trevr}2 for a single band on a single processor {for one timestep.}.  
These tests were run in serial to eliminate the complications of load balancing in measuring the expected algorithmic scaling.  The $N$ shown on the x-axis is the total number of particles participating in that physics step, so it is larger for gravity ($N=N_\textrm{gas}+N_\textrm{star}+N_\textrm{dark matter}$) than for RT, where $N=N_\textrm{gas}+N_\textrm{star}$.  
The clusters of blue points with similar run times to the gravity red crosses represent results for the same run.  Thus {\sc trevr}2 RT is similar in wallclock cost to gravity in {\sc gasoline}2. These figures have many points, representing runs with every combination of reduced particle numbers for each type independently ($\times 1$, $\times 1/10$ and $\times 1/100$).  This demonstrates that the RT scaling is with respect to the total $N$, without regard to whether they are sources or absorbers.

The parallel implementation of {\sc trevr2} is very similar to the gravity code and the parallel performance is also similar.  In parallel on 320 cores, one global calculation ($N_\textrm{active}=N$)
of SPH forces takes 0.6 s, gravity takes 4.3 s and RT takes 3.4 s for {\sc trevr}2 with a single band on this galaxy with all particles.  The original {\sc trevr} takes 45 s without refinement and 93 s with modest refinement for a HI band ($\tau_\textrm{refine}=1$).  
{The key point is that the general behaviour that radiation and gravity take similar time is maintained.  We intend to optimize and test the parallel version in future work, using a more modern code base such as {\sc changa} \citep{jetleyEt2008}.}

Figure~\ref{fig:scalingactive} shows that the code achieves close to $N_\textrm{active} \log_2 N$ scaling on small time steps.   The overall speedup is limited by the tree build cost to a factor $\sim 30$ in serial.  
In parallel, the tree build work is well balanced so the benefit is improved ($\gtrsim 40$ on 320 cores).  

The {\sc trevr2} algorithm performs a lot of work that is unrelated to the specific band, including tree walking and cell intersections.  Thus adding extra bands is very cheap, as it is a matter of an inner loop over bands with no branching, which is easily optimized by compilers.  The red points in the figure~\ref{fig:scalingactive} were measured for 10 bands in the inner loop which increases the run time by a factor of only $\sim 1.3$!
We note that this was performed as an experiment, and the multiband version of {\sc trevr2} is still being developed.

\subsection{Str\"{o}mgren Tests}\label{sec:strom}

{These tests repeat the exact tests applied to {\sc trevr} in \cite{grond2019}, following} \citet{ilievEt06}, to evolve the Str\"{o}mgren sphere around a source emitting $5 \times 10^{48}$ photons per second in a medium with $n_H = 10^{-3}$ cm$^{-3}$. 
There are two versions, one isothermal ($10^4$ K, monchromatic at 13.6 eV with $\sigma=6.3\times 10^{-18}$ cm$^2${\bf, called test1 in Iliev et al.}) and another with Hydrogen-based cooling (nominally with $\sigma=1.63 \times 10^{-18}$ cm$^{2}$ and fixed heating per absorption{\bf, called test2}).  We note that different codes in the Iliev comparison had different physics (e.g. minimum allowed ionization fractions) and RT algorithms, including multiple bands in a few cases.  We refer the reader to \citet{ilievEt06} and \cite{grond2019} for additional details.  {Our tests use purely hydrogen gas with case B recombination, hydrogen line cooling and a single band as given in \citealt{grond2019}.  The results can be directly compared to those for {\sc rsph}, {\sc traphic} and {\sc trevr}}.

We show HI fraction slices for {\sc trevr2} in figure~\ref{fig:stromiso} at 30 and 500 Myr for the isothermal case.  This plot was notably used in \cite{pawlikSchaye08} to assess the noise and direction accuracy of {\sc traphic} (their figure 6 for 500 Myr).  {\sc trevr2} is smoother than standard {\sc traphic} but not as spherical as the original {\sc trevr} for {\bf$N_\textrm{ray}=48$}, due to the angular discretization of {\sc healpix}.  

In figure~\ref{fig:stromrad}, we show {\sc trevr2} results for the Str\"{o}mgren test with cooling, as well as results from the original figure in \cite{ilievEt06}. 
We note that the original {\sc trevr} results had larger radii for lower resolution (\citealp{grond2019}, section 3.3.3), which can be attributed to the lack of the optically thick correction from section~\ref{sec:optthick}.  
We include curves for {\sc trevr2} with both the standard transmission averaging (red) and linear absorption averaging (black).  
Temperature profiles at the same times are {shown} in figure~\ref{fig:stromtemp}.

The absorption average result is too small at intermediate times.   
This issue arises because rays from the source often traverse just the inner part of cells with a large radial extent.  
A linearly averaged absorption strongly reflects material properties at a greater radius than the intersecting segment in cases where the interior is low opacity (ionized in this case) and the exterior is high opacity.  As neutral hydrogen cross-sections are high, even a small neutral volume can dominate the average.  {\sc trevr2} with transmission averaging is not systematically affected by the inclusion of higher opacity gas in a small part of the tree cells.  We note in passing that we explored using piece-wise linear absorption in each cell with interpolation, and this is also able to correct for the issue in a test with a simple gradient like this (see section~\ref{sec:discussion} for further discussion).

\begin{figure*}
    \centering
	 \includegraphics[width=\textwidth]{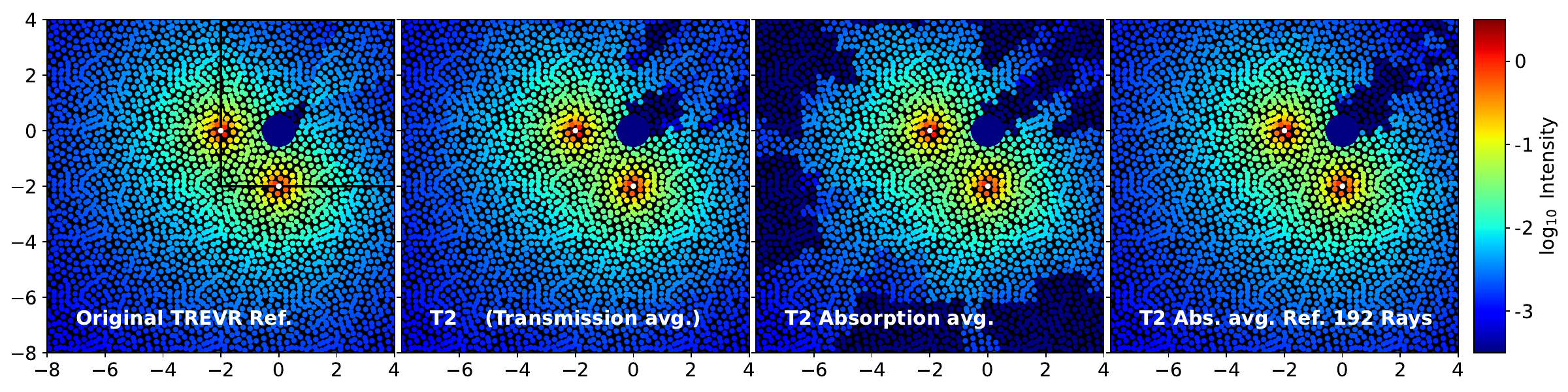}
    \caption{Cloud irradiated by two sources.  Individual particle radiation energy densities are {shown via colours indicated in a log$_{10}$ scale on the colour bar}.  From the left: {\sc trevr} solution with refinement accurately reproduces the exact solution.  The black square marks the \citet{wunsch2021} box boundaries.  The standard {\sc trevr2} solution (with transmission averaging) is qualitatively similar to the accurate solution, with erroneous shadowed areas related to the healpix beam sizes.  Photons escape to infinity in all directions.  The {\sc trevr2} solution with linearly averaged cell absorption experiences catastrophic flux loss once the sources and cloud are combined into one cell (receiving gas at $x\lesssim -6$ or $y \lesssim -6$). The {\sc trevr2} solution with absorption averaging, refinement and $N_\textrm{ray}=192$ is closer to the original {\sc trevr} (but expensive).}
    \label{fig:TRF13_4panel}
\end{figure*}

\subsection{Cloud irradiated by two sources}\label{sec:treerayfig13}

This test is designed to be equivalent to the test with the same name presented in \citet{wunsch2021}.  Their test set-up consists of a low density, low opacity medium with a 20 $M_\odot$ sphere with radius 0.5 pc at the origin and two sources, luminosity $L=0.5$ each, one at $x=-2$ pc and one at $y=-2$ pc and run for one step.  Temperatures are not specified.  However, assuming a cool neutral medium that is 75 \% Hydrogen, the optical depth at 13.6 eV across the diameter of the sphere is 22700 ($\alpha=22700$ pc$^{-1}$).  From the exact solution presented (\citealp{wunsch2021}, figure~13), the low density medium is essentially transparent $\alpha \ll 1$.   We use a simplified version of this test with a single wavelength and fixed $\alpha = 10000$ inside the sphere and $\alpha = 10^{-4}$ elsewhere but otherwise geometrically identical.   The SPH low density particles are $0.25$ pc apart in a glass, while the high density sphere has a $0.0625$ pc spacing as is apparent in figure~\ref{fig:TRF13_4panel}.

The original test from \cite{wunsch2021} is well designed because the sources and cloud will be combined into a single cell that is much smaller than the entire volume, so that cell opening systematics are readily apparent.  In appendix~\ref{appendix:tree}, we illustrate how poor test design can hide systematics on tests like this.  In our initial conditions, we place the centre of the simulation volume at (-2.1,-2.1,-2.1) to allow the sources and cloud to be combined into a cell in a similar way to \cite{wunsch2021}.

We draw attention to the panel labelled (a) in figure 13 from \cite{wunsch2021}, with a resolution of 0.047 per finest cell (about 5 times better than our resolution outside the sphere).  Here the radiation energy catastrophically drops to zero 3 pc from the dense sphere.  
This radius is due to their Barnes-Hut opening angle being set to $\theta_\textrm{lim} = 1.0$ allowing a single cell containing both sources and the sphere to be used.  The size of the cell is $\sim3$ pc on each side.  Assuming it includes the whole sphere this gives an average $\alpha \sim 440$.  The optical depths are thus $\tau \sim 1320$ and no radiation escapes beyond $3$ pc from the sphere.  This behaviour is {\it normal}, all tree-based approaches will routinely use combined cells.  For other cases shown in \cite{wunsch2021}, $\theta_\textrm{lim} \leq 0.5$, so that this problem radius is pushed to $6$ pc or more, just outside the test simulation domain.  The critical point is that the problem is not resolved.  There are many astrophysical scenarios with sources and absorbers within a pc of each other (e.g. Giant Molecular Clouds, AGN in Galaxies) and a simulation domain large enough so that the problem distance is inside.  We contend that decreasing the opening angle further is not a practical solution for dynamic ranges of many orders of magnitude, as it reverts it back to a $\bigO{N^2}$-type scheme.  In practice, there are often multiple, distributed sources so the problem will not be as sharply obvious as in a test with just two adjacent sources.

Figure~\ref{fig:TRF13_4panel} shows slices through the various {\sc trevr2} solutions that illustrate different strategies to address systematics associated with linear averaging of cell absorption. 
The original {\sc trevr} method can use refinement ($\tau_\textrm{refine} = 0.01$) and get very close to the exact solution modulo the glass particle distribution (leftmost panel). It is comparable to the panel labelled 'analytical' in \citet{wunsch2021}.  The opacity contrast here is so sharp that any optical depth refinement criterion gives similar results (e.g. $\tau_\textrm{refine} \lesssim 1$).  The solution region here is larger than the original \citeauthor{wunsch2021} box (shown in the first panel), to demonstrate the problems with cell merging at larger radii.

{\sc trevr2} with transmission averaging (second panel) captures the essential features of the exact solution with a small volume lost to erroneous shadows associated with the finite beam sizes.  Importantly, this approach maintains the $N \log_2 N$ scaling of the method.  The third panel shows that linear absorption averaging experiences catastrophic failure once combined cells are allowed by the opening criterion.  On the figure this occurs for receiving gas at 
$x\lesssim -6$ or $y \lesssim -6$ ($\sim 4$ away from the combined cell edge in all directions).

{\sc trevr} automatically uses smaller cells (narrower beams) at the cost of an $\bigO{N^{4/3}}$ expense. In the rightmost panel,  {\sc trevr2} must increase the number of beams (192 {\sc healpix} rays shown) to benefit from such refinement and still has rough shadows near the cloud.  The refinement expense is similar to {\sc trevr}, with a less accurate result and thus this is not an effective strategy for {\sc trevr2}.   This can be compared to panel (c) in figure 13 in \citet{wunsch2021} which also uses 192 rays.

We give a more quantitative assessment of these results in table~\ref{tab:escape}.  
The first section of the table shows escape fractions for the two sources plus cloud test (measured on a sphere at $r=6$).  
The expected result is very close to the refined {\sc trevr} answer: around 98 \% of the radiation should escape.  We define the covering factor in terms of the fraction of the sphere (solid angle) where the flux is reduced to half or less its value by absorption.  
By these measures {\sc trevr2} is able to give qualitatively good answers without sacrificing efficiency via the transmission average.

\begin{figure*}
    \centering
    \includegraphics[width=\textwidth]{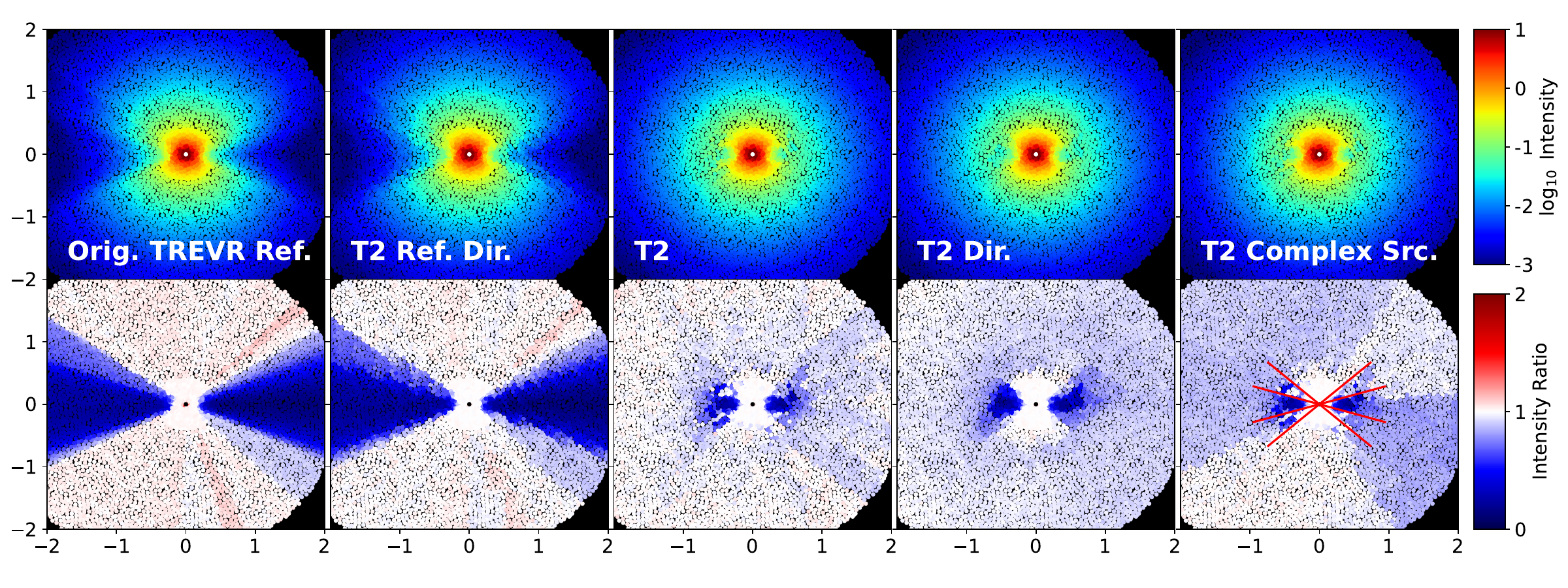}
	 \caption{Top row: Log$_{10}$ Intensity for a source flanked by two clouds. Bottom row:  Intensity relative to exact answer without clouds ($4 \pi r^2 e^r\ \times$ intensity), to show {shadows and} asymmetry due to clouds (linear scale). From the left, the first column shows original {\sc trevr} with aggressive refinement which is qualitatively correct with some errors at the shadow edges.   
    {\sc trevr}2 with refinement and adjusting ray directions to point directly at source rather than along healpix beam centre is also accurate.  The middle column shows standard {\sc trevr}2 (without refinement).  The clouds are diluted via cell mergers until the shadows fade at larger radii.  The 4th panel is {\sc trevr2} with rays directed at the source in place of the {\sc healpix} direction. The rightmost panel shows the complex source model overlaid with the 8 healpix directions closest to the slice plane in red.  The black corner areas have no particles.}
    \label{fig:twoclouds}
\end{figure*}

\begin{figure*}
    \centering
	    \includegraphics[width=\textwidth]{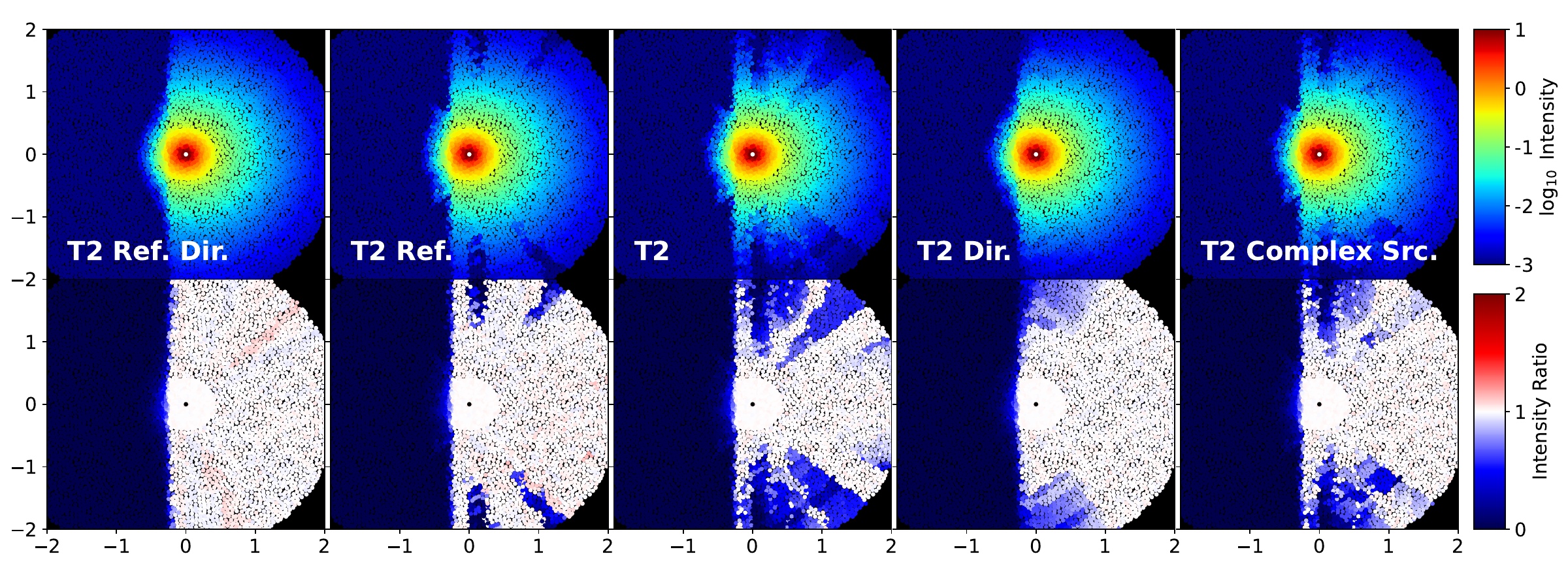}
    \caption{Top row: Log$_{10}$ Intensity for a source next to a dense wall.  Bottom row: {Intensity relative to exact answer without wall ($4 \pi r^2 e^r\ \times$ intensity)}, to show asymmetry due to wall (linear scale).   From the left:  {\sc Trevr2} with refinement and rays directed at the source is very accurate.  {\sc trevr2} with refinement has spurious shadow streams.  Standard {\sc trevr2} has merged cells causing problems at larger distances. {\sc trevr2} with rays directed at the source is better.  The last panel shows the complex source model.}
    \label{fig:wall}
\end{figure*}

\begin{table}
 \caption{Quantitative assessment of test results.  This summarizes three different tests using escape fraction as a figure of merit compared to the exact result.  The escape fraction is the flux fraction reaching the outer radius from the centre.  The covering factor is the fraction of the solid angle blocked by asymmetric features (clouds or walls).  Blocked means less than half the flux seen without those features arrives at the outer radius.  The outer radius is 6 for the first (two sources) test and 2 otherwise. }
 \label{tab:escape}
 \begin{tabular}{llll}
  \hline
  Test & Method & Escape & Covering\\
   \  & \  & Fraction & Factor\\
   \hline
     Two Sources & Exact (no cloud) & 100 \% & 0 \\
  + Cloud & {\sc trevr} Refined & 97.6 \% & 0.036 \\
         \  & {\sc trevr}2  & 91.7 \% & 0.013 \\
    \  & {\sc trevr}2 Absorption average     & 33.7 \% & 0.684\\
       \  & {\sc trevr}2 Abs. avg. Ref. 192 Rays & 96.5 \% & 0.035 \\      \  & {\sc trevr}2 Complex Source & 93.6 \% & 0.018 \\
  \hline
  Source & Exact (no clouds) & 13.53 \% & 0 \\
  + Two clouds & Exact & 12.81 \% & 0.069 \\
   \  & {\sc trevr} Refined & 12.97 \% & 0.057 \\
      \  & {\sc trevr}2 Ref. Directed & 12.75 \% & 0.059 \\
         \  & {\sc trevr}2 Refined & 13.31  \% & 0.021 \\
                   \  & {\sc trevr}2 Directed     & 12.88 \% & 0.000 \\
          \  & {\sc trevr}2      & 13.39 \% & 0.000 \\
          \  & {\sc trevr}2 Absorption Average & 13.29 \% & 0.000 \\      
 \  & {\sc trevr}2 Complex Source &     12.60 \% & 0.000  \\
  \hline
    Source & Exact (no wall) & 13.53 \% & 0 \\
    + Wall & Exact & 7.61 \% & 0.44 \\
   \  & {\sc trevr} Refined & 7.62 \% & 0.45 \\
   \  & {\sc trevr}2 Ref. Directed & 7.50 \% & 0.45 \\
      \  & {\sc trevr}2 Refined & 6.73 \% & 0.51 \\
         \  & {\sc trevr}2 Directed & 6.16 \% & 0.52 \\
         \  & {\sc trevr}2 & 5.42 \% & 0.59 \\
 \  & {\sc trevr}2 Absorption Average & 4.60 \% & 0.67  \\
     \  & {\sc trevr}2 Complex Source & 6.12 \% & 0.55 \\
   \hline      
 \end{tabular}
\end{table}

\subsection{Complex Source Tests}\label{sec:complex}

These tests illustrate the challenging nature of complex sources where the opacity changes dramatically as a function of direction.  Typically, the solution is fairly good close to the source, where the opening criterion separately opens cells containing dense gas and sources.  As shown below, this performance can typically be maintained through the use of absorption-based refinement ($\tau_\textrm{refine} = 0.01$).   Refinement is not prohibitively expensive with a handful of sources (such as in these tests).  However, it is not a viable option for galaxy-type simulations with many sources.  We note that ray tracing methods are extremely accurate for spherically symmetric tests with smooth absorption profiles, so we don't include such tests here and focus on more difficult tests.

\subsubsection{Two clouds and one source}\label{sec:twoclouds}

Figure~\ref{fig:twoclouds} shows results for a source flanked by two moderately dense clouds.  The medium has $\alpha=1$ except in the radius=$0.1$ clouds which have $\alpha=10$.  The simulation volume is a sphere out to 2.5 centred on the origin.  The source {has $L=1$ and} is at $(0.17,0.17,0.17)$ and the clouds are offset by $\pm 0.25$ along the diagonal,  which means they are in the top right corner of the entire simulations volume.  This choice prevents symmetric placement of tree boundaries helping maintain symmetry in the results.  In particular, if objects are placed in different octants of the simulation volume, there is a large tree boundary (side length of the entire simulation) which prevents the opening criterion from ever combining them into one cell (illustrated further in appendix~\ref{appendix:tree}).  
The particle spacing is 0.0625 in the outer part and 0.0209 within 0.4 of the source.  This ensures the clouds are well resolved.

In addition $(1,1,1)/\sqrt{3}$ is not a {\sc healpix} direction or aligned with any cell edge, so there the results do not artificially look better and the angular features represent typically behaviour for the method.  
A key point regarding figure~\ref{fig:twoclouds} is that the horizontal axis is along the diagonal $(1,1,1)/\sqrt{3}$ with zero at the source, so that the exact answer should show the shadows falling in this direction.  The vertical axis is a direction perpendicular to the first axis $(1,-1,0)/\sqrt{2}$. 

Compared to the previous test, a crucial difference is that the shadows are darker but not empty of radiation and extend to infinity in the absence of scattering.  The top row in each case shows the mean intensity in a slice.  The bottom row is the same simulation {intensity data relative to the exact} solution without the clouds, so as to focus on the quality of the shadows {caused by the clouds} and the effect of different methods.  Here the answer should be 1 where the clouds are having no effect.  In general, it should always be 1 or less.

{\sc trevr} and {\sc trevr2} can maintain qualitatively good answers via refinement (as shown in the first two columns of figure~\ref{fig:twoclouds} respectively).  For this test we have focused on separating out the effects of refinement and directional corrections in {\sc trevr2}.  

{\sc trevr2} with the standard opening criterion without refinement (i.e. that maintains the fast $N \log_2 N$ property) is shown in the middle column.  The simulation switches to using a combined source with moderate absorption ($\tau \sim 1$) beyond $r \sim 1$.  Unlike the previous test, this does not wipe out the flux.  The correct answer is a lower flux in the shadows, which cover about 7 \% of the solid angle, as viewed by the source.

Qualitatively, we can say that most of the radiation that should escape to $r=2$ still does in all cases, as shown in the second section of table~\ref{tab:escape}.   The accurate answer is 12.8 \% reduced from 13.5 \% ($e^{-2}$) due to the presence of the clouds. However, {\sc trevr2} gives 13.4 \% once the source is merged with the clouds at $r \gtrsim 1$.  In this case, the linear absorption average and transmission average give similar results, as the optical depths per cell are small.

We note that {\sc trevr2} is typically less accurate than {\sc trevr}, even for simple cases with one source, because the beam is constrained to point in a {\sc healpix} direction rather than directly at the source.  
For a simple test like this, {\sc trevr2} can be temporarily modified to point at the source (here the origin) rather than along the centre of the {\sc healpix} beam for a given ray.  This is shown in the table with the 'Directed' cases, which give substantially improved escape percentages. 

The fourth column shows that directing the rays at the source is quantitatively better with less noise, but still results in the angular structure being lost at larger radii.  This is comes from the use of a combined cell that cannot represent the angular structure and ultimately fails to recover the covering factor.  This is an intrinsic limitation of the tree-based method without aggressive refinement of some form.  

Correcting the direction is easy for a single source test.  It might be generalized to point at a weighted centre of flux or luminosity, except for the problem of some sources being heavily obscured, which could not be known in advance without multiple ray casts.   This presents an avenue for future work, to see if the concept can be generalized.

The final column shows the complex source model result, with detailed remarks in section~\ref{sec:discussion}.  In this case, the optical depth from absorbers near the source is estimated at the tree build stage.  This leads to artifacts associated with specific healpix directions seen in the lower panel.  It should be noted that the source is also offset so that the radiation tree is does not have reflection symmetries in this projection.  Thus oppositely directed healpix direction do not give the same result.   The overlaid healpix beam directions give a sense of the angular resolution.

\subsubsection{A wall and one source}\label{sec:wall}

Figure~\ref{fig:wall} shows results for a source offset from a dense wall of material ($\alpha=10$), where the remainder is low opacity ($\alpha=1$).  The particle spacing and offset are the same as in section~\ref{sec:twoclouds} (source {\bf$L=1$}, offset by $(0.17,0.17,0.17)$ ) to avoid helpful symmetries and the wall is offset along the main diagonal -0.25 from the source.  This setup is reminiscent of a blister HII region.  It is a useful test because radiation can escape effectively over roughly half the sphere.  We present quantitative results in the final section of table~\ref{tab:escape}.  As above, the horizontal axis is along the diagonal $(1,1,1)/\sqrt{3}$, so that the wall is vertical in figure~\ref{fig:wall}.

We show top and bottom rows with the full intensity and the modification due to the wall respectively, as in the prior section.  As in other tests, {\sc trevr2} with refinement is quantitatively very good if rays are directed at the source (leftmost panels).  {\sc trevr2} is also fairly good with refinement alone (second column).  

The middle column shows standard, fast {\sc trevr2}. The result is qualitatively acceptable but a bit low for the escape fraction.  The healpix directions lead to picking up extra absorption from the wall near the source, which takes the form of shadow streaks in the second and third columns.  The fourth column 
shows that even standard, fast {\sc trevr2} benefits from the directional fix.   The directional fix substantially improves the escape and covering factor values, as shown in the table.  The primary remaining source of error is when a cell combines the source with absorbing wall material via the opening angle for $r \gtrsim 1$ near the y-axis.   The rightmost column shows the complex source model result and is discussed further in section~\ref{sec:discussion}.

The wall feature has a fairly substantial optical depth ($\tau \sim 20$ at $r=2$).  All the {\sc trevr2} results discussed so far use the standard transmission average.  Conversely, {\sc trevr2} with linear absorption average gives a terrible result on this test as shown in the table (but no figure panels).   

\section{Discussion}\label{sec:discussion}

An overall goal for this work was to evaluate the trade-offs associated with fast reverse ray tracing.  It is relatively easy to get good results on spherical test cases (e.g. section~\ref{sec:strom}).  We were particularly interested in harder test problems without any helpful symmetries, either in the test itself or in the way it is realized as a simulation, that would make the results look artificially good (section~\ref{sec:complex}).   The test results demonstrate that {\sc trevr2} functions well, with qualitatively good results. In particular, if we focus on the average intensity (for heating and chemistry), rather than detailed shadows, the results are quantitatively good as well, as long as we use a transmission average for higher optical depth cases.  

We note that the errors tend to take the form of shadows in the wrong place at large radii rather than excess flux, as shown in the bottom row results in figures~\ref{fig:twoclouds} and~\ref{fig:wall}.  The tests we employed typically have one or two sources for the entire volume.   In more realistic situations, there are often many sources and possibly a uniform background as well.  In particular, the {\sc trevr2} results are always good next to a source.  The problems begin once source and absorber cells are combined at larger radii and it is reasonable to expect that other sources would become important at these distances.  Another important factor, to be addressed properly in future work, is that scattering and equivalent processes can redirect light to fill in shadows and potentially facilitate radiation escape from complex structures.

However, we would still like to improve the basic results for complex arrangements of sources and absorbers.  Specifically, standard {\sc trevr2} results become less accurate with respect to angular distributions in such cases. This is due to a fundamental tension between the complex arrangement of these sources (which we can address at cost via refinement with {\sc trevr}) and maintaining the speed of the method by combining cells to retain $N \log_2 N$ scaling.  

It is worth pointing out that the cell opening criterion nominally limits errors associated with multiple sources by limiting the ratio of the higher order moments to the monopole associated with the combined source.  This applies for gravity and RT without absorption.   However, with absorption, a combined cell source location can be moved entirely outside or inside the main absorbers in the cell, becoming much brighter or much dimmer,  respectively.  In this case, the opening criterion is only a weak constraint on the accuracy.

We investigated several additional strategies to get better results on these cases.  For example, one can use a linear fit to the opacity in the combined cells.  This only works well when there is a simple, large-scale opacity gradient as is the case for the Str\"omgren test (section~\ref{sec:strom}).
In general, it is quite common to have the absorption go up and back down again within a single large cell.  We explored this further by implementing and testing a piece-wise constant and linear fit.  However, as should be expected, this cannot adapt to an arbitrary distribution of opacity.  More specifically, it did not substantially improve the results for the complex sources tests of section~\ref{sec:complex}.   This is because, at large distances from the source, it will be treated as a single cell where the absorbers are centrally-peaked features inside a large, otherwise more transparent volume.  In this case, the fact that the source has preferred escape directions (e.g. along the vertical direction in figure~\ref{fig:twoclouds}), is impossible to capture via a simple fitting function.

Another strategy we tested was to embed an angle-based {\it complex source model} for the absorption around the cell source in the source cell data structure, as described in section~\ref{sec:complexsourcemodel}.  We show the visual model results in the rightmost columns of figures~\ref{fig:twoclouds} and~\ref{fig:wall}.  
For the two cloud test shown in figure~\ref{fig:twoclouds}, the model restores the existence of the shadows to some degree, but their angular distribution is too wide and somewhat misdirected compared to the more accurate results in the leftmost column of figure~\ref{fig:twoclouds}.  Similarly, for the wall test shown in figure~\ref{fig:wall}, the complex source only partly removes the spurious shadows associated with combined cells at larger radii.  The crude shadows are a result of the ray optical depth combination process which can only take into account two cells at a time and is unable to account for structure within those cells.  In this sense the complex source model is not equivalent to refinement.

We have included complex source model results in table~\ref{tab:escape} with the label 'Complex Source'.  It can be seen that the numerical answers for escape fraction and covering factor are only moderately improved for the two source plus cloud, high opacity test. We note that most photons escape in this test, so the differences are not very significant either way.  Visually it is practically the same as the standard {\sc trevr2} panels in figure~\ref{fig:TRF13_4panel}.  

For the source and two clouds test, however, the quantitative answers are substantially improved with the complex source model.  This indicates that while the complex source model only marginally improves the angular treatment, the average absorption is more representative.  In particular, the escape fraction result is significantly improved.   The result for the wall test is also quantitatively better with an escape fraction that is closer to the exact answer.  However, the errors are still significant.

As noted previously, the tree build is 40 \% more expensive with the complex source model and the node storage is significantly larger.  The overall runtime is similar, as there is slightly less RT work overall when $N_\textrm{active} \sim N$.   However, in production runs with a wide range of time steps, we expect to benefit a lot from the $N_\textrm{active} \log_2 N$ scaling and a more expensive tree build limits those benefits.   In practice, the complex source model is also fairly complicated to implement.   We conclude that the complex source model is only somewhat useful in its current form.  However, it is an avenue for continued exploration, particularly more complicated and accurate versions of the model.

Another potential strategy is to apply refinement in a more limited way.  We have shown that refinement works well for mixed sources and absorbers, but it can dramatically increase the runtime.  This is because the optical depth refinement scheme outlined in \cite{grond2019} is triggered throughout the volume for the high optical depths associated with the ionization cross-section of neutral hydrogen and helium in galaxy simulations.    

Refinement also causes mismatches in the angular resolution that negatively affects {\sc trevr2} compared to the original {\sc trevr}.   For simple tests (sec~\ref{sec:complex}), we showed that modifying the rays to point at the source rather than the standard {\sc healpix} directions can help. This provides some basic angular adaptivity, but it is not clear how to do it generally, for many sources in a beam.   Related strategies could include avoiding refinement for distant or heavily obscured sources, which commonly occur in astrophysics.  Such strategies would likely require multiple traces per ray while iteratively refining the absorption, and potentially also the angle (spawning further sub-rays), once it is established that a source will make an important contribution.  We note that {\sc healpix} is well-suited to adaptivity in  principle, though it would make the code much more complex in practice.

\section{Conclusions}\label{sec:con}

In this work, we described {\sc trevr2}, a fast update to the {\sc trevr} algorithm for reverse ray tracing.  The new algorithm is designed to efficiently estimate mean intensities (to calculate heating and chemistry), at the expense of angular accuracy except in relatively simple scenarios (e.g. spherical symmetry).  The method was implemented in the {\sc gasoline2} parallel code.  It was demonstrated to achieve the target $N_\textrm{active} \log_2 N$ scaling in practice, taking similar runtime to gravity on a standard zoomed galaxy.  In particular, the radiative transfer work scales with the number of active particles and thus retains the dramatic speed-up of per-particle timesteps relative to other radiative transfer methods. 

Particularly useful is the low overhead for many radiation bands.  Many bands do not dramatically change the cost versus a single band.  We will explore ways to model multiple bands and their impact on galaxy environment in a companion paper (Baumschlager et al., in prep).

The new method includes several key improvements over {\sc trevr} and similar reverse ray tracing algorithms (e.g. {\sc treeray}). The most important is transmission averaging rather than linear absorption averaging.  This is essential for high optical depths or the absorption is likely to be severely over-estimated for combined cells.  We demonstrate this through both standard (e.g. Str\"omgren sphere) and novel test problems with different, relatively geometrically simple combinations of sources and absorbers.

These tests demonstrate a key tension for tree-based reverse ray tracing.  Sources and strong absorbers are often close together and will be combined into a single cell by the standard tree-opening criteria responsible for the $N_\textrm{active} \log_2 N$ scaling.  We can recover accurate results via refinement (losing the fast scaling) or accept some qualitative and quantitative reductions in accuracy (e.g. inaccurate angular distributions).    

We outline various strategies to improve the accuracy and their performance costs.  This includes an optional complex source model, a potential way forward that does not change the scaling.  Optimal strategies to combine features such as transmission averaging with different approaches to beam averaging (e.g. that used in {\sc treeray}), refinement and other aspects of {\sc healpix}-based schemes will require future work.  

Another consideration for future work is optimally combining chemistry and RT calculations.  {\sc treeray} applies an iterative approach to ensuring both opacities and radiation fields are representative.  
The {\sc trevr2} strategy is to scale with the active subset, so that it is efficient to adapt time steps where chemistry is changing.  However, with radiation this is not necessarily a local effect.  Changing opacities may effect radiation fields across the simulation, in principle.  Thus, timesteps may need to be modified over large areas when such events occur.  
Demonstrating the effectiveness of the {\sc trevr2} time step strategy when solving detailed chemistry (i.e. whether it can avoid large increases computational cost) is a key goal for future work.  

An additional, complicating aspect of real astrophysical systems is scattering and recombination processes that effectively make all gas a source of radiation.  The tests shown here indicate {\sc trevr2} will run efficiently with all gas as a source.  The accuracy remains to be assessed.  Of particular interest is the fact that these processes will cause diffusion in the radiation direction that reduces the importance of shadowing and potentially increasing the escape of photons from complex source structures (previously highlighted by \citealt{hasegawa2010}).  We expect this will increase the usefulness of methods like {\sc trevr2} and we intend to include these processes in in future work.

We are particularly interested in finding strategies that work well in galaxy simulations, where the optical depths can be very high and sources strongly clustered within dense gas. 
A related, difficult issue is that galaxy simulations cannot resolve the dense substructure around star clusters and other important radiation sources.  Thus, unresolved complex source and absorber distributions will need to be addressed in future work by all simulation groups.   It is likely that treatments will need to be somewhat statistical and allow for unresolved sources that have preferred escape directions.  

Overall, it seems clear that RT strategies must be adapted for the astrophysical systems of interest and adjusted for the current simulation resolution.   In addition, figures of merit must be chosen carefully and it is unreasonable to expect that a single simulation is representative.  For example, radiation escape from galaxies is likely to be highly stochastic, even before we consider the nature of unresolved sources.  Repeated simulations will be needed to establish statistically robust results (see e.g. \citealt{kellerchaos}).  This highlights the importance of fast RT methods, to explore variance.



\section*{Acknowledgements}

The authors would like to thank Ben Keller, Padraic Odesse and the anonymous referee for helpful comments.  
JW is supported by a Discovery Grant from NSERC of Canada.
BB and SS acknowledge support from the Research Council of Norway through NFR Young Research Talents Grant 276043.  Computational resources for this project were enabled by Compute Canada/Digital Alliance Canada and parallel simulations were carried out on the Niagara computing cluster.

\section*{Data Availability}

The simulations described here were performed with an updated version of the public {\sc gasoline2} simulation code.  Updated code will be made available via existing online repositories.   Simulation data and initial conditions are available on request.



\bibliographystyle{mnras}
\bibliography{references} 



\appendix

\counterwithin{figure}{section}

\section{Avoiding bias in RT Tests}\label{appendix:tree}
\begin{figure}
    \centering
	 \includegraphics[width=\columnwidth]{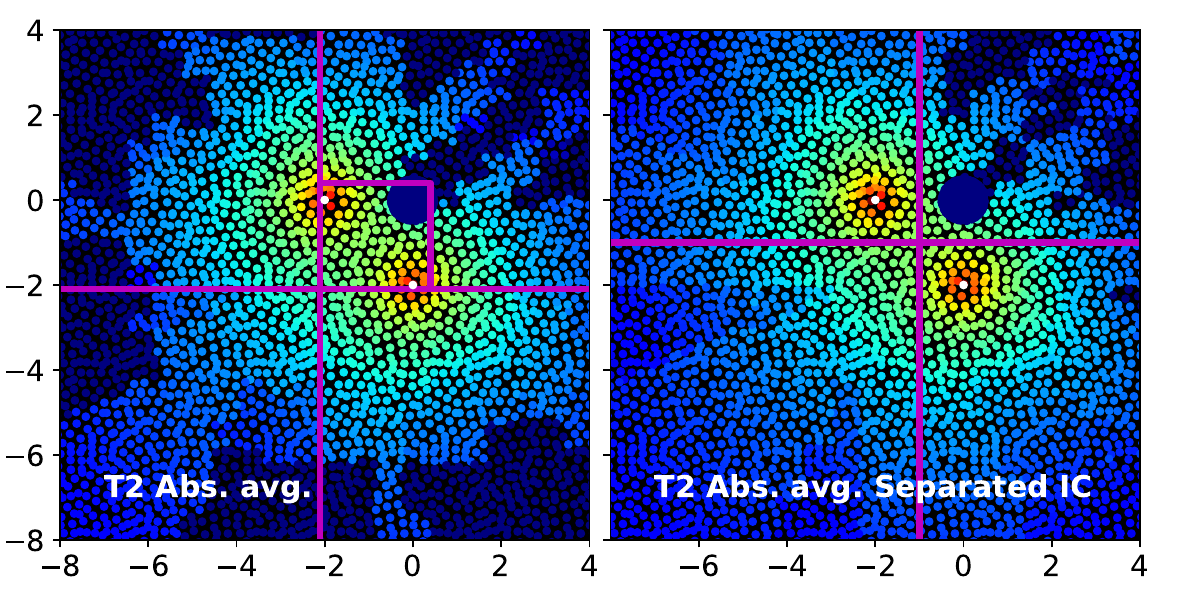}
    \caption{Cloud irradiated by two sources with absorption averaging, showing dependence on initial condition design.  The left panel (a) shows our standard initial condition where the cloud and sources are offset from the domain centre and the first cell that contains all three is indicated.  In the right panel (b), the centre of the simulation is instead between the components so that they are in separate octants.  The only cell that contains all three is the whole simulation which will never be used for RT.  This artificially improves the result as shown. }
    \label{fig:treeeffect}
\end{figure}

For the tests discussed in section~\ref{sec:tests}, it is important not to let the special geometry of a test problem create misleading results.  This is illustrated in figure~\ref{fig:treeeffect}.  In practice, objects of interest are unlikely to be arranged neatly around the exact centre of the simulation.  However, with test problems this is a common choice and it can easily lead to objects being assigned to different octants of the simulation.  In this case, the only cell to contain all the objects is the root cell that is the whole simulation.  This cell is always opened by the opening criterion so that the sources never merge, with artificially good results.

For the left panel in the figure, which is a case commonly encountered in practice, sources and dense objects are combined in a moderately-sized cell that the RT will use, producing systematic effects if they exist (such as for absorption averaging used for these figures).  In the right panel, the first cell to hold all three objects is the root cell, which is always opened and systematics are avoided by construction.  In all the tests in this work, we always place the objects of interest in the same octant to provide a fair test.


\bsp	
\label{lastpage}
\end{document}